\documentclass[floatfix,groupedaddress,superscriptaddress,aps,showpacs,amsmath,amssymb,prl,
longbibliography,
twocolumn,a4paper,10pt]{revtex4-2}
\usepackage{lmodern}
\usepackage{bm} 
\usepackage{caption}
\usepackage{newtxmath}
\usepackage{siunitx}
\usepackage{amsbsy}
\usepackage{braket}
\usepackage{graphicx}
\usepackage{subcaption}
\usepackage{placeins}

\usepackage{ragged2e}  
\usepackage{microtype} 

\usepackage{caption}
\usepackage{booktabs}
\usepackage{dcolumn}
\usepackage{comment}
\usepackage{xcolor}
\usepackage{soul} 
\usepackage[colorlinks,
linkcolor=blue,
anchorcolor=blue,
citecolor=blue,
filecolor=blue,
menucolor=blue,
runcolor=blue,
urlcolor=blue,
frenchlinks=true]{hyperref}

\usepackage[noabbrev, nameinlink]{cleveref}
\begin{document}

\title{Efficient Implementation of a Single-Qutrit Gate Set via Coherent Control}

\author{Xiang-Min Yu}
\thanks{These authors contributed equally.}
\affiliation{National Laboratory of Solid State Microstructures, School of Physics, Nanjing University, Nanjing 210093, China}
\affiliation{Shishan Laboratory, Nanjing University, Suzhou 215163, China}
\affiliation{Jiangsu Key Laboratory of Quantum Information Science and Technology, Nanjing University, Suzhou 215163, China}
\affiliation{Synergetic Innovation Center of Quantum Information and Quantum Physics, University of Science and Technology of China, Hefei, Anhui 230026, China}
\affiliation{Hefei National Laboratory, Hefei 230088, China}

\author{Xiang Deng}
\thanks{These authors contributed equally.}
\affiliation{National Laboratory of Solid State Microstructures, School of Physics, Nanjing University, Nanjing 210093, China}
\affiliation{Shishan Laboratory, Nanjing University, Suzhou 215163, China}
\affiliation{Jiangsu Key Laboratory of Quantum Information Science and Technology, Nanjing University, Suzhou 215163, China}

\author{Wen Zheng}
\email{zhengwen@nju.edu.cn}
\affiliation{National Laboratory of Solid State Microstructures, School of Physics, Nanjing University, Nanjing 210093, China}
\affiliation{Shishan Laboratory, Nanjing University, Suzhou 215163, China}
\affiliation{Jiangsu Key Laboratory of Quantum Information Science and Technology, Nanjing University, Suzhou 215163, China}

\author{Wei Xin}
\thanks{These authors contributed equally.}
\affiliation{National Laboratory of Solid State Microstructures, School of Physics, Nanjing University, Nanjing 210093, China}
\affiliation{Shishan Laboratory, Nanjing University, Suzhou 215163, China}
\affiliation{Jiangsu Key Laboratory of Quantum Information Science and Technology, Nanjing University, Suzhou 215163, China}

\author{Tao Zhang}
\affiliation{National Laboratory of Solid State Microstructures, School of Physics, Nanjing University, Nanjing 210093, China}
\affiliation{Shishan Laboratory, Nanjing University, Suzhou 215163, China}
\affiliation{Jiangsu Key Laboratory of Quantum Information Science and Technology, Nanjing University, Suzhou 215163, China}
\author{Hanxin Che}
\affiliation{National Laboratory of Solid State Microstructures, School of Physics, Nanjing University, Nanjing 210093, China}
\affiliation{Shishan Laboratory, Nanjing University, Suzhou 215163, China}
\affiliation{Jiangsu Key Laboratory of Quantum Information Science and Technology, Nanjing University, Suzhou 215163, China}
\author{Kun Zhou}
\affiliation{National Laboratory of Solid State Microstructures, School of Physics, Nanjing University, Nanjing 210093, China}
\affiliation{Shishan Laboratory, Nanjing University, Suzhou 215163, China}
\affiliation{Jiangsu Key Laboratory of Quantum Information Science and Technology, Nanjing University, Suzhou 215163, China}
\affiliation{Synergetic Innovation Center of Quantum Information and Quantum Physics, University of Science and Technology of China, Hefei, Anhui 230026, China}
\affiliation{Hefei National Laboratory, Hefei 230088, China}
\author{Haoyu Zhou}
\affiliation{National Laboratory of Solid State Microstructures, School of Physics, Nanjing University, Nanjing 210093, China}
\affiliation{Shishan Laboratory, Nanjing University, Suzhou 215163, China}
\affiliation{Jiangsu Key Laboratory of Quantum Information Science and Technology, Nanjing University, Suzhou 215163, China}
\author{Yangyang Ge}
\affiliation{National Laboratory of Solid State Microstructures, School of Physics, Nanjing University, Nanjing 210093, China}
\affiliation{Shishan Laboratory, Nanjing University, Suzhou 215163, China}
\affiliation{Jiangsu Key Laboratory of Quantum Information Science and Technology, Nanjing University, Suzhou 215163, China}
\author{Zhenchuan Zhang}
\affiliation{Shishan Laboratory, Nanjing University, Suzhou 215163, China}
\affiliation{Jiangsu Key Laboratory of Quantum Information Science and Technology, Nanjing University, Suzhou 215163, China}
\author{Wanli Huang}
\affiliation{Shishan Laboratory, Nanjing University, Suzhou 215163, China}
\affiliation{Jiangsu Key Laboratory of Quantum Information Science and Technology, Nanjing University, Suzhou 215163, China}
\author{Haoyang Cai}
\affiliation{National Laboratory of Solid State Microstructures, School of Physics, Nanjing University, Nanjing 210093, China}
\affiliation{Shishan Laboratory, Nanjing University, Suzhou 215163, China}
\affiliation{Jiangsu Key Laboratory of Quantum Information Science and Technology, Nanjing University, Suzhou 215163, China}
\author{Xianke Li}
\affiliation{National Laboratory of Solid State Microstructures, School of Physics, Nanjing University, Nanjing 210093, China}
\affiliation{Shishan Laboratory, Nanjing University, Suzhou 215163, China}
\affiliation{Jiangsu Key Laboratory of Quantum Information Science and Technology, Nanjing University, Suzhou 215163, China}

\author{Jie Zhao}
\affiliation{National Laboratory of Solid State Microstructures, School of Physics, Nanjing University, Nanjing 210093, China}
\affiliation{Shishan Laboratory, Nanjing University, Suzhou 215163, China}
\affiliation{Jiangsu Key Laboratory of Quantum Information Science and Technology, Nanjing University, Suzhou 215163, China}
\author{Xinsheng Tan}
\affiliation{National Laboratory of Solid State Microstructures, School of Physics, Nanjing University, Nanjing 210093, China}
\affiliation{Shishan Laboratory, Nanjing University, Suzhou 215163, China}
\affiliation{Jiangsu Key Laboratory of Quantum Information Science and Technology, Nanjing University, Suzhou 215163, China}
\affiliation{Synergetic Innovation Center of Quantum Information and Quantum Physics, University of Science and Technology of China, Hefei, Anhui 230026, China}
\affiliation{Hefei National Laboratory, Hefei 230088, China}
\author{Yu Zhang}
\affiliation{National Laboratory of Solid State Microstructures, School of Physics, Nanjing University, Nanjing 210093, China}
\affiliation{Shishan Laboratory, Nanjing University, Suzhou 215163, China}
\affiliation{Jiangsu Key Laboratory of Quantum Information Science and Technology, Nanjing University, Suzhou 215163, China}

\author{Shao-Xiong Li}
\email{shaoxiong.li@nju.edu.cn}
\affiliation{National Laboratory of Solid State Microstructures, School of Physics, Nanjing University, Nanjing 210093, China}
\affiliation{Shishan Laboratory, Nanjing University, Suzhou 215163, China}
\affiliation{Jiangsu Key Laboratory of Quantum Information Science and Technology, Nanjing University, Suzhou 215163, China}
\affiliation{Synergetic Innovation Center of Quantum Information and Quantum Physics, University of Science and Technology of China, Hefei, Anhui 230026, China}
\affiliation{Hefei National Laboratory, Hefei 230088, China}
\author{Yang Yu}
\email{yuyang@nju.edu.cn}
\affiliation{National Laboratory of Solid State Microstructures, School of Physics, Nanjing University, Nanjing 210093, China}
\affiliation{Shishan Laboratory, Nanjing University, Suzhou 215163, China}
\affiliation{Jiangsu Key Laboratory of Quantum Information Science and Technology, Nanjing University, Suzhou 215163, China}
\affiliation{Synergetic Innovation Center of Quantum Information and Quantum Physics, University of Science and Technology of China, Hefei, Anhui 230026, China}
\affiliation{Hefei National Laboratory, Hefei 230088, China}

\date{\today}

\begin{abstract}
Qutrits offer the potential for enhanced quantum computation by exploiting an enlarged Hilbert space.
However,
the synthesis of high-fidelity and fast qutrit gates, particularly for single qutrits, remains an ongoing challenge, as it involves overcoming intrinsic constraints in quantum platforms.
Here, we develop a novel framework for the efficient implementation of a single-qutrit gate set via coherent control, leveraging SU(3) dynamics while obviating platform-specific constraints such as those arising from the selection rule.
As a proof-of-principle demonstration, we realize 35-ns qutrit Hadamard and X gates using a superconducting transmon, achieving an average fidelity of 99.5\%, as verified by randomized benchmarking.
We further demonstrate two paradigmatic quantum circuits, which can be naturally extended to scalable qudit algorithms for phase estimation and parity check.
In addition, we propose an SU(3)-based decomposition strategy for an arbitrary single-qutrit gate and numerically demonstrate its substantial efficiency improvement over conventional SU(2)-based protocols.
By addressing the challenge of efficiently implementing single-qutrit gates, our protocol paves the way for realizing high-performance qutrit processors in diverse quantum platforms.
\end{abstract}
                             
\maketitle

\textit{\textcolor{blue}{Introduction}}---%
Encoding quantum information in a larger-dimensional Hilbert space can significantly reduce circuit depth and improve algorithmic efficiency \cite{nielsen2010,gokhale2020extending,wang2020qudits,kiktenko2025}.
Theoretical studies have suggested that qutrit-based computation offers notable advantages in quantum error correction (QEC) due to the additional level enabling compact encoding schemes, richer syndrome information, and improved error thresholds \cite{bullock2007,campbell2014enhanced,majumdar2018quantum,prakash2020magic,schmidt2024error,ma2023}.
Driven by these prospects, experimental research on qutrit computation is rapidly advancing across various platforms, including trapped ions \cite{mc2005trapped,ringbauer2022universal,edmunds2025,nikolaeva2025}, superconducting circuits \cite{morvan2021qutrit,blok2021,roy2023two,liu2023performing,goss2024,nguyen2024,tripathi2025,champion2025}, and others \cite{fu2022,chi2022programmable,lindon2023complete,zhou2024,guo2024}.
Specifically, using a superconducting qutrit-based architecture, QEC surpassing the break-even point has been demonstrated \cite{brock2025quantum}, marking a milestone toward fault-tolerant quantum computing.

Notable efforts have been made to the implementation of two-qutrit gate operations through straightforward extensions of existing two-qubit schemes \cite{goss2022high,luo2023experimental},
making significant progress toward advancing programmable qutrit-based processors.
As another fundamental component in quantum circuits, single-qutrit gates are considerably more prevalent than their two-qutrit counterparts \cite{roy2023two}.
However, single-qutrit operations are governed by the SU(3) group, which, in contrast to SU(2) operations for 
qubits, is generated by Gell-Mann matrices \cite{liu2023performing}.
The non-trivial commutation relationships among the matrices pose a substantial obstacle to the efficient implementation of single-qutrit gates.

For example, the widely employed scheme for the single-qutrit gate synthesis involves gate decomposition \cite{divincenzo2000,liu2023performing,fischer2023universal}, which 
entails decomposing a qutrit gate into a sequence of SU(2) operations restricted to two-level subspaces.
However, this approach necessitates extended gate sequences, which inevitably lead
to increased operation times and considerable error accumulation, ultimately undermining the potential advantages of qutrit-based architectures \cite{wack2021}.
Alternatively, optimal control techniques \cite{wu2020high,cho2024direct} have been applied to design hardware-specific control pulses capable of performing the desired qutrit gates, yet this method consumes substantial computational resources and requires extensive calibrations tailored to specific hardware architectures \cite{cho2024direct}, thus limiting its scalability in practical processors.
Therefore, there is a pressing need for the development of an efficient qutrit gate synthesis strategy, akin to those established for 
qubits, to mitigate the trade-offs inherent in existing schemes.

\begin{figure*}
  \centering
  \includegraphics[width=17cm]{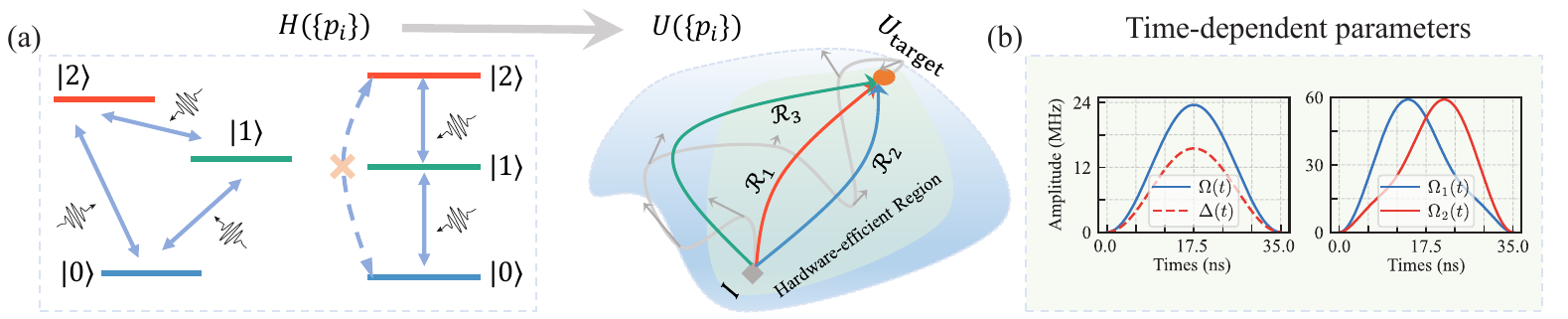}
\caption{ \justifying 
(a) 
Typical $\Delta$- and $\Xi$-type qutrit configurations. While the prevalent $\Xi$-type Hamiltonian is constrained by selection rules that complicate gate synthesis, 
our protocol simultaneously applies two coherent pulses to efficiently realize $SU(3)$ gates in one step.
By analytically solving $U(\{p_i\}) = U_{\mathrm{target}}$ for a parameterized Hamiltonian $H(\{p_i\})$, we identify multiple evolution paths (e.g., $\mathcal{R}_1$-$\mathcal{R}_3$) from the identity $I$ to the target unitary within the parameter space $\{p_i\}$.
This contrasts with optimal control (gray trace), which relies on iterative search and may exceed the experimentally feasible region (green area). 
(b) As a proof-of-principle demonstration, we identify a set of time-dependent parameters to efficiently implement the primitive  H and X gates. Note that for the H gate, $ \Omega_1(t) = \Omega_2(t) = \Omega(t)$, while for the X gate, $\Omega_1(t) = \Omega_2(T-t)$ and $\Delta(t) = 0$.
  \label{fig:1} 
}
\end{figure*}

In this Letter, 
we develop a novel protocol for the efficient implementation of a single-qutrit gate set by coherently manipulating SU(3) dynamics.
Analogous to their qubit counterparts, we introduce qutrit Hadamard (H) and X gates serving as primitive operations in qutrit-based quantum computation \cite{wang2020qudits}.
Both gates are demonstrated on a transmon \cite{Koch2007}, a fabrication-friendly superconducting circuit \cite{you2011,gu2017,krantz2019,Blais2021}, achieving an average fidelity of 99.5\% with a gate time of 35 ns.
To explore the capability of the protocol, we implement two key quantum circuits. The first is qutrit Ramsey interferometry \cite{shlyakhov2018quantum,suppmaterials} analogous to a triple-slit experiment, which 
holds potential for high-precision phase estimation and intrinsically outperforms
its qubit counterparts. The second is the parity-check algorithm \cite{gedik2015computational,suppmaterials}, achieving a special quantum speedup without requiring entanglement resources.
Finally, we present a universal SU(3) decomposition strategy and demonstrate its high compilation efficiency through numerical analysis.

\textit{\textcolor{blue}{Theoretical Framework}}---%
Considering the flexibility of SU(3) dynamics,
we implement the single-qutrit Clifford group $\mathcal{C}_{3}$ using generators $\{\mathrm{H, S, X, Z}\}$ \cite{ringbauer2022universal, liu2023performing}, which are respectively defined as
\begin{equation} \label{eq1}
\begin{aligned}
\mathrm{H} &= \frac{1}{\sqrt{3}} \sum_{j,k=0}^{2} \omega^{jk} \ket{k}\bra{j},
\quad \mathrm{X} = \sum_{j=0}^{2} \ket{j \oplus 1}\bra{j},
\\
\mathrm{Z} &= \sum_{j=0}^{2} \omega^{j}\ket{j}\bra{j}, \quad
\mathrm{S} = \sum_{j=0}^{2} \omega^{j(j+1)/2}\ket{j}\bra{j},
\end{aligned}
\end{equation}
where $\omega = e^{2\pi i/3}$ is the primitive cube root of unity and $\oplus$ denotes addition modulo~3. 
Compared to SU(2)-based schemes, synthesizing the 216 distinct single-qutrit Clifford gates using the generating set defined by Eq.~\eqref{eq1} reduces the average native-gate cost from 5.25 to 1.667~\cite{liu2023performing}, corresponding to a 68.2\% improvement in compilation efficiency.
When supplemented with a non-Clifford gate such as $\mathrm{T} = \mathrm{diag}(1, e^{2\pi i/9}, e^{-2\pi i/9})$ \cite{howard2012qudit}, the Clifford generators defined by Eq.~(\ref{eq1}) extend to a universal single-qutrit gate set \cite{ringbauer2022universal,prakash2018normal}.
Since the phase gates Z, S, and T can be implemented using extended virtual phase control, where each phase gate is realized by updating the reference phase of all subsequent microwave pulses \cite{lindon2023complete,mckay2017efficient},
our efforts focus on realizing the H and X gates in a single step via bichromatic drives that coherently manipulate the three-level dynamics.

As illustrated in Fig.~\ref{fig:1}(a), $\Xi$-type qutrit platforms, which are prevalent in natural and artificial systems, are constrained by a selection rule that forbids the $\ket{0} \leftrightarrow \ket{2}$ transition.
A previously explored $\Delta$-type system
employs highly connected couplings to implement the H gate with three delicate drives, resulting in a zero gauge-invariant phase \cite{yurtalan2020implementation}, and the X gate via three equal-strength drives, yielding a $\pi/2$ phase \cite{tao2021experimental}.
However, scaling these systems requires extreme anharmonicity or computational-basis rearrangement, posing significant challenges in prevalent platforms.
We thus directly propose a protocol for $\Xi$-type systems that identifies efficient paths to target gates, bypassing the typical iterative difficulties encountered in optimal control.
Specifically, our approach leverages analytical solutions of a qutrit driven by two coherent pulses in the two-photon resonant regime.
The resulting effective Hamiltonian is given by
\begin{equation} \label{eq2}
H/\hbar = \Delta \ket{1} \bra{1} +  \frac{1}{2} (\Omega_{1} \ket{0} \bra{1} + \Omega_{2} \ket{1} \bra{2} + h.c.),
\end{equation} 
where $\Delta$ denotes the single-photon detuning, and $\Omega_{1,2}$ are the Rabi frequencies of coherent drives for the transitions $\ket{0} \leftrightarrow \ket{1} $ and $\ket{1} \leftrightarrow \ket{2} $, respectively.

First, we consider a simple case where $\Omega_{1,2}$ and $\Delta$ remain constant over time, in which case the time-evolution operator can be formulated as the exponential map of the time-independent Hamiltonian
\begin{equation} \label{eq:exp}
U(T) = \exp(-\frac{i}{\hbar} H T) = \sum_{i,j = 0}^{2} U_{ij} \ket{i} \bra{j},
\end{equation}
where the exact form of each matrix element $U_{ij}$ ($i,j = 0,1,2$) can be derived analytically via the SU(3) Lie algebra relations \cite{zhang2022robust,zhang2024high}. For a given evolution period $T$, 
there always exist appropriate choices of $\Omega_{1,2}$ and $\Delta$ such that
the outcome $|U_{ij}| = 1/\sqrt{3}$ is achieved for all indices $i$ and $j$. In this scenario, the qutrit H gate can be represented as the time-evolution operator $U(T)$ sandwiched between two phase gates, $U_{d_{1}}$ and  $U_{d_{2}}$, i.e. $\mathrm{H} = U_{d_{1}} \cdot U(T) \cdot U_{d_{2}}$ \cite{shlyakhov2018quantum}. 
However, for practical applications, it is essential that the control pulses start and end at zero amplitude, ensuring compatibility with experimental hardware.
Note that the analytical framework of Eq.~(\ref{eq:exp}) remains valid for time-dependent pulses provided that $\Omega_{1,2}(t)$ and   $\Delta(t)$ share a common temporal envelope.
We therefore suggest the widely used phase-modulated chirped pulses to implement the qutrit H gate,
where $\Omega_{1,2}(t)$ and $\Delta(t)$ feature cosine-shaped envelopes, as depicted in the left panel of Fig.~\ref{fig:1}(b); see the Appendix A for details.

The qutrit X gate exhibits the well-known chiral 
property of
population transfer along the cyclic pathway $\ket{0} \to \ket{1} \to \ket{2} \to \ket{0}$, indicating that its physical implementation requires the breaking of time-reversal symmetry (TRS) \cite{roushan2017chiral,shapira2023quantum}.
Since such real-valued time-independent Hamiltonians inherently preserve TRS,  
they are insufficient for implementing the X gate.
We then turn to the time-dependent case.
For simplicity, we focus on the single-photon resonant regime, viz. $\Delta = 0$, in which the corresponding dynamical Lewis-Riesenfeld (LR) invariant $I(t)$ \cite{lewis1969,chen2011lewis}, satisfying $ \partial I(t) / \partial t + i/\hbar [ H(t), I(t) ] = 0 $, has previously been extensively discussed in state preparation \cite{chen2012engineering,GueryOdelin2019,yu2025quantum}.
Here, we leverage the LR invariant theory to explicitly derive the qutrit time-evolution operator:
\begin{equation} \label{eq:inv}
U(T) = \sum_{j=0,\pm}e^{i\theta_{j}(T)} \ket{\phi_{j}(T)} \bra{\phi_{j}(0)},
\end{equation}
with the LR phase $\theta_{j}(T)$ defined as
$
\theta_{j}(T) = \frac{1}{\hbar} \int_{0}^{T}\left\langle \phi_{j}(t) \right| i\hbar \frac{\partial}{\partial t}- H(t) \left| \phi_{j}(t) \right\rangle dt.
$
Here, $\ket{\phi_{j}(t)}$ is the instantaneous eigenstate of the dynamical invariant $I(t)$, satisfying $I(t) \ket{\phi_{j}(t)} = \lambda _{j} \ket{\phi_{j}(t)}$, 
with $\lambda_j$ being its time-independent eigenvalue.
To explicitly construct the X gate, we impose appropriate boundary conditions on the invariant eigenstates and employ a polynomial ansatz with tunable auxiliary parameters to precisely target the required LR phases. This analytical procedure allows us to derive the specific time-dependent Rabi frequencies, as shown in the right panel of Fig.~\ref{fig:1}(b).
A detailed derivation is provided in Appendix B.

Taken together, the protocol enables an efficient one-step implementation of qutrit H and X gates via direct SU(3) dynamical control.
A similar procedure holds promise for the synthesis of other qutrit gates.
Further details of the pulse design for qutrit gates are available in Supplemental Material \cite{suppmaterials}.

\begin{figure}
    \centering
    \includegraphics[width=0.480\textwidth, height=0.354\textheight]{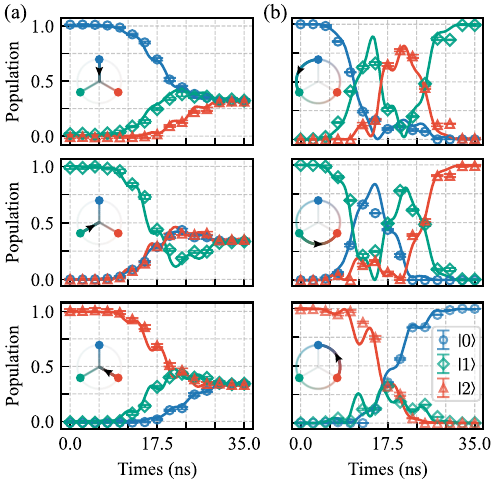}
     \caption{\justifying 
     Population dynamics for initial states $|0\rangle$, $|1\rangle$, and $|2\rangle$ under (a) the H gate, evolving toward their respective equally weighted superpositions, and (b) the X gate, exhibiting chiral population transfer.
     The population dynamics are indicated by corresponding inset symbols.
     Experimental data (dots) with standard deviation (error bars) are compared to theoretical simulations (solid lines).
     \label{fig:2}
     }
\end{figure}

\textit{\textcolor{blue}{Experimental Results}}---%
To experimentally demonstrate our framework, we implement the protocol on a transmon, exemplifying its broad applicability without loss of generality.
While the anharmonicity of 
$-$193 MHz enables selective addressing of individual transitions, it also faces significant crosstalk, with control pulses inevitably driving off-resonantly non-target transitions, leading to substantial coherent errors \cite{li2024universal}. 
To mitigate these effects, we develop a systematic calibration procedure to 
optimize the drive parameters \cite{suppmaterials}.

We begin by characterizing the qutrit gates through an investigation of population dynamics.
Fig.~\ref{fig:2}(a) presents the results for the H gate, where the transmon is initialized in states $\ket{0}$, $\ket{1}$, and $\ket{2}$, shown from top to bottom, respectively. Each initialization is followed by 
35-ns control pulses
tailored for the H gate.
The Hamiltonian
Eq. 
~\eqref{eq2} exhibits a parity-like symmetry $PH(t)P=H(t)$, which arises from the condition $\Omega_1(t) = \Omega_2(t)$ with $P = \ket{0}\bra{2} + \ket{2}\bra{0}$, resulting in mirrored dynamics.
Similarly, for the X gate, as shown in Fig.~\ref{fig:2}(b), the distinct chiral 
property
revealed by population transfer
is evident.
Importantly,
the $\mathrm{X}^{-1}$ gate can be directly implemented by reversing the control pulses in time.
In this case, 
Eq. 
~\eqref{eq2} possesses a symmetry $PH(t)P=H(T-t)$, assuming $\Omega_1(t) = \Omega_2(T-t)$ and $\Delta = 0$,  thereby yielding $\mathcal{P}\mathcal{T}$-symmetry dynamics.
The results indicate the high-fidelity implementation of both H and X gates, demonstrating efficient and coherent control over the qutrit dynamics.

\begin{figure}
    \centering
    \includegraphics[width=7.5cm]{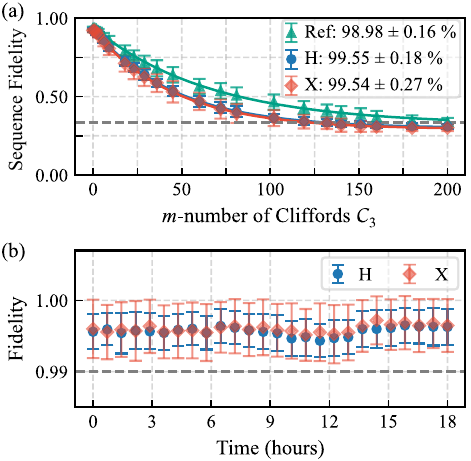}
    \caption{\justifying 
    (a) The sequence fidelity, 
    the measured population in the $\ket{0}$ state, as a function of the length of the Clifford-random gate sequence for both standard RB (green), and IRB with the H gate (blue) and X gate (orange) interleaved.
    The ground state population converges to the dashed reference line at 1/3, indicating that the qutrit state, after undergoing randomly sampled Clifford-based circuits, indeed tends toward the maximally mixed state.
    (b) The 
    fidelities of qutrit H and X gates extracted from repeated IRB measurements over continuous 18-hour period with all parameters kept consistent.
    The dashed reference line 
    indicates
    the 99\% gate fidelity threshold.
    }
    \label{fig:3} 
\end{figure}

Routinely, to assess the gate performance, we employ randomized benchmarking (RB), a widely used technique that estimates the average gate error independent of state preparation and measurement errors \cite{morvan2021qutrit, kononenko2021characterization}.
As shown in Fig. \ref{fig:3}(a), we perform the standard 
RB by repeatedly applying gate sequences that are randomly sampled from the single-qutrit Clifford group $\mathcal{C}_3$. 
Due to the inherent properties of the Clifford group, averaging over these randomly sampled sequences will lead to a depolarizing channel.
The resulting exponential decay of the population allows us to extract the depolarization parameter $p_{c}$.
From this, the average error per Clifford gate is $r_{c}=0.0102$, given by $r_{c} = (1-p_{c})(1-1/d)$, where $d=3$ represents the dimension of the qutrit Hilbert space and $p_c$ is the depolarization parameter extracted from exponential fit.

To characterize specific Clifford gates, especially the H and X gates of interest, we adopt interleaved RB (IRB), in which the target gate is interleaved after each Clifford gate in the standard RB sequence.
The gate error can be determined by comparing the depolarization parameters obtained with and without the interleaved gate, then given by $r_{g}=(1-p_{g}/p_{c})(1-1/d)$, where $p_{g}$ is the depolarization parameter obtained from the IRB.
From the experimental results, we find the individual average gate errors for the H gate $r_{\mathrm{H}}=0.0045$ and for the X gate $r_{\mathrm{X}}=0.0046$.
Notably, the extracted average errors are consistent with the decoherence-induced error limit, suggesting that coherent errors have been effectively suppressed through our calibration procedure \cite{suppmaterials}.
Moreover, to further characterize the long-time stability of our experimental setup, we monitor the average gate error by repeatedly performing IRB over an 18-hour period without any recalibration.
As shown in Fig.~\ref{fig:3}(b), the gate errors remain stable for each gate throughout the monitoring duration, confirming that our protocol can be reliably executed over prolonged periods on current devices.

\begin{figure}
    \centering
    \includegraphics[width=7.5cm]{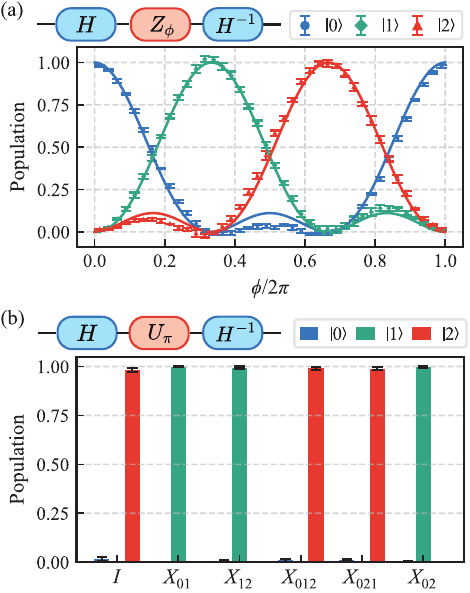}
    \caption{\justifying 
    (a) Qutrit Ramsey interferometry. A qutrit initialized in the state $\ket{0}$ undergoes a Ramsey sequence: an H gate, a phase gate $Z_{\phi} = \mathrm{diag} (1, \exp(\phi), \exp(2\phi) )$, and an inverse H gate. The final  populations in each basis are measured as a function of the delayed phase $\phi$. The dots represent experimental results, while the associated solid lines reflect theoretical predictions. (b) Parity-check algorithm. The qutrit is first initialized in the state $\ket{1}$, followed by an H gate, a permutation operation $\pi$ implemented via an X-type qutrit 
    gate $U_{\pi}$, and an inverse H gate. The final populations in each basis are measured as a function of the permutation. Depending on the permutation parity, the qutrit deterministically evolves into either $\ket{1}$ (even) or $\ket{2}$ (odd).}
    \label{fig:4}
\end{figure}

\begin{figure}
  \centering
  \includegraphics[width=8cm]{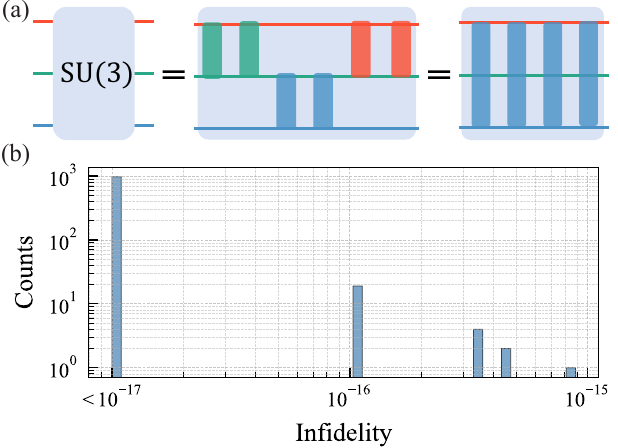}
\caption{ \justifying 
(a) For an SU(3) gate, conventional SU(2)-based decomposition requires six native gates (green, blue, and red boxes), whereas our strategy requires only four (blue boxes). (b) Validation of the proposed strategy through the successful compilation of 1,000 random SU(3) gates.
}
  \label{fig:5} 
\end{figure}

Finally, we demonstrate the applications of our protocol by executing two paradigmatic quantum circuits, which can be extended to single-qudit scenarios.
All involved qutrit gates are directly implemented in one step via coherent control, thereby showcasing the advantage of our protocol.
The first, as shown in Fig.~\ref{fig:4}(a), is qutrit Ramsey interferometry, an approach widely used in quantum metrology \cite{shlyakhov2018quantum,suppmaterials}.
Theoretically, the final populations $P_{j}$ of the qutrit quantum 
state $\ket{j}$ ($j=0, 1, 2$), depend on the delayed phase $\phi$, and can be expressed as
\begin{equation}
P_{j}=\frac{1}{9} [1 + 2\cos(\phi - 2\pi j/3)]^{2}.
\end{equation}
The experimental results exhibit overall agreement with theoretical predictions.
Combined with the Kitaev algorithm \cite{danilin2018quantum}, the iterative qutrit Ramsey 
sequences enable
high-precision phase extraction.
In particular, as detailed in Supplemental Material \cite{suppmaterials}, we show that the probability density for obtaining a phase estimate $\tilde{\phi}$ after $N$ iterations is given by:
\begin{align}
p(\tilde{\phi}|\phi ) &\propto \frac{1}{2\pi d^{N}}  \frac{\sin^{2}(d^{N}  \delta \phi /2)}{\sin^{2}( \delta \phi /2)}.
\end{align}
As the number of iterations $N$ or the dimension $d$ increases, the distribution becomes progressively sharper, with the width of the main peak scaling as $1/d^{N}$.
Consequently, the qudit Ramsey interferometry allows the extraction of more information per iteration, reducing the number of individual measurements required to achieve a given degree of precision by a factor of $O(\log{d})$.

The other is the well-known parity-check algorithm designed to determine whether a given permutation within the dihedral group is even or odd.
For an initial state $\ket{m}$ satisfying $\gcd(m,d)=1$, the algorithm yield orthogonal final states, depending on the permutation parity:
\begin{equation}
\ket{\psi_{f}} = 
\begin{cases} 
\ket{m} & \text{for even permutation,}  \\ 
\ket{d-m} & \text{for odd permutation.} 
\end{cases}
\end{equation}
As shown in Fig.~\ref{fig:4}(b),
a qutrit initialized in the state $\ket{1}$ will deterministically evolve into the state $\ket{1}$ for even permutations and into the state $\ket{2}$ for odd permutations.
Notably, 
classical algorithms require at least two queries to determine the parity of a permutation, that is one to identify the cyclic shift and another to determine its direction. In contrast, the quantum algorithm enables parity determination with a single query, achieving a two-to-one speedup without requiring entanglement \cite{gedik2015computational}.

\textit{\textcolor{blue}{Universal $SU(3)$ Synthesis}}---Beyond Clifford gates, our native qutrit gates enable efficient universal SU(3) compilation.
Specifically, for an arbitrary operation $U \in SU(3)$, we propose the decomposition strategy:
\begin{equation} \label{Eq:decom}
U = \mathrm{Z}_{\phi_1} \mathrm{H Z}_{\phi_2} \mathrm{H Z}_{\phi_3} \mathrm{H Z}_{\phi_4} \mathrm{H Z}_{\phi_5},
\end{equation}
where $\mathrm{Z}_{\phi_j} = \mathrm{diag} (1, e^{i\phi^{(1)}_j}, e^{i(\phi^{(1)}_j + \phi^{(2)}_j) })$ is the generalized phase gate, realized virtually at the software level.

The strategy provides clear advantages over the SU(2)-based decomposition. 
While SU(2)-based schemes typically require 6 native Hadamard gates \cite{lindon2023complete,liu2023performing,li2024universal},
our strategy reduces the required native gate count by 1/3, as illustrated in Fig.~\ref{fig:5}(a).
Although a rigorous mathematical proof of Eq.~\eqref{Eq:decom} remains an open problem, Fig.~\ref{fig:5}(b) confirms the universality and high fidelity of this strategy, evidenced by numerically testing on 1,000 random SU(3) gates, following the methods established in prior work \cite{divincenzo2000}. In Appendix C, we systematically summarize the advantages of our protocols over conventional SU(2)-based schemes across three key areas: arbitrary SU(3) compilation, Clifford group synthesis, and specific H-gate implementation.

\textit{\textcolor{blue}{Conclusions}}---In summary, we present a simple, efficient one-step protocol for implementing a single-qutrit gate set by directly manipulating the three-dimensional Hilbert space.
As a proof-of-principle demonstration, we analytically design H and X gate pulses within 35 ns, experimentally achieving a reliable fidelity of 99.5\%. We examine the advantages of our protocol by applying it to two application-relevant quantum circuits, validating its feasibility for qutrit computation. 
Furthermore, we theoretically explore the distinct potential of qutrits for high-precision quantum metrology.
In particular, building upon our experimentally demonstrated native qutrit gates, we propose a universal SU(3) decomposition strategy, offering enhanced compilation efficiency as confirmed by numerical analysis.
Our platform-independent protocol can serve as a universal toolbox for controlling widespread three-level systems and lays the foundation for the systematic design of qudit gates across diverse quantum hardware.
~\\

\textit{Acknowledgments}---%
We are grateful to the anonymous referees for their valuable guidance; their suggestions were instrumental in improving the quality of this paper.
We thank Wei Fang, Jiayu Ding and Orkesh Nurbolat for technical support.
This work was partially supported by
the Quantum Science and Technology-National Science and Technology Major Project (Grant Nos. 2021ZD0301702, 2024ZD0302000),
Natural Science Foundation of Jiangsu Province (Grant No. BK20232002),
National Natural Science Foundation of China (Grant Nos. U21A20436 and 12074179),
and
Natural Science Foundation of Shandong Province (Grant No. ZR2023LZH002).

~\\

\textit{Data availability}---%
The data that support the findings of this Letter are not publicly available.
The data are available from the authors upon reasonable request.

%

\onecolumngrid 
\vspace{2em} 
\begin{center}
    \textbf{End Matter} 
\end{center}
\vspace{1em} 
\twocolumngrid 

\appendix

\textit{Appendix A: Qutrit H gate}---%
The exact form of each matrix element $U_{ij}$ ($i,j = 0,1,2$) for Eq.~\eqref{eq:exp} can be explicitly expressed as \cite{zhang2022robust,zhang2024high}: 
\begin{equation} \label{eqs3}
\begin{aligned}
U_{00} &= \cos^{2} \theta + e^{-i \delta} ( \cos \frac{A}{2} + i \frac{\Delta}{\Omega} \sin \frac{A}{2} ) \sin^{2} \theta \\
U_{11} &= e^{-i \delta} (\cos \frac{A}{2} - i \frac{\Delta}{\Omega} \sin \frac{A}{2} )  \\
U_{22} &= \sin^{2} \theta + e^{-i \delta} ( \cos \frac{A}{2} + i \frac{\Delta}{\Omega} \sin \frac{A}{2} ) \cos^{2} \theta  \\
U_{01} &= U_{10} = -i e^{-i \delta} \frac{\Omega_{0}}{\Omega} \sin \theta \sin \frac{A}{2}  \\
U_{12} &= U_{21} = -i e^{-i \delta} \frac{\Omega_{0}}{\Omega} \cos \theta \sin \frac{A}{2}  \\
U_{02} &= U_{20} = e^{-i \delta} ( \cos \frac{A}{2} + i \frac{\Delta}{\Omega} \sin \frac{A}{2} - e^{i \delta} ) \sin \theta \cos \theta
\end{aligned}
\end{equation} 
where the reduced parameters $\theta$, $\Omega_{0}$, $\Omega$, $A$ and $\delta$ are defined as $\tan \theta = \Omega_{1}/\Omega_{2}$, $\Omega_{0} = \sqrt{\Omega_{1}^{2} + \Omega_{2}^{2}}$, $\Omega = \sqrt{\Omega_{0}^{2} + \Delta^{2}}$, $A = \Omega T$ and $\delta = \Delta T/2$, respectively. Any $3 \times 3$ unitary operator with equal-modulus matrix elements, i.e., $|U_{ij}| = 1 / \sqrt{3}$, can be represented as either a qutrit H gate or its inverse, sandwiched between two diagonal phase gates \cite{shlyakhov2018quantum}. 
To minimize the required Rabi frequencies, we select the optimal solution defined by $|A| = 4.0410$ and $|\delta| = 0.8525$, which yields the integrated pulse parameters $\Omega_{1,2}T = 2.5906$ and $|\Delta T| = 1.7050$.
For a gate duration of $T = 35$ ns, this yields $\Omega_{1,2}/2\pi = 11.7801$ MHz and $\Delta/2\pi = 7.7531$ MHz. The qutrit H gate is thus explicitly constructed as:
\begin{equation}  \label{eqs5}
\mathrm{H} = \begin{pmatrix}
 1 & 0 & 0 \\
 0 & e^{i(\frac{2\pi}{3} + \delta  )}  & 0 \\
 0 & 0 & e^{-i\frac{2\pi}{3}}
\end{pmatrix}
\cdot U(T) \cdot
\begin{pmatrix}
 e^{-i\frac{\pi}{6}} & 0 & 0 \\
 0 & e^{i(\frac{\pi}{2} + \delta  )}  & 0 \\
 0 & 0 & e^{-i\frac{5\pi}{6}}
\end{pmatrix}.
\end{equation} 

While the analytical derivation assumes square pulses, experimental implementation requires zero-amplitude boundary conditions. Consequently, we employ phase-modulated chirped pulses where $\Omega_{1,2}(t)$ and $\Delta(t)$ share a common temporal profile. To realize the H gate, the integrated pulse areas must satisfy $\int_{0}^{T}\Omega_{1,2} (t)dt = 2.5906$ and $\int_{0}^{T} \Delta(t) dt = 1.7050$. The corresponding laboratory-frame driving fields are given by:
\begin{equation}\label{eqs6}
\begin{aligned}
\tilde{\Omega}_{1}(t) &=  \Omega_{1}(t) \cos\left(\omega _{01}t  + \int_{0}^{t} \Delta (\tau)d\tau\right),  \\
\tilde{\Omega}_{2}(t) &=  \Omega_{2}(t) \cos\left(\omega _{12}t  - \int_{0}^{t} \Delta (\tau)d\tau\right). 
\end{aligned}
\end{equation}
where $\omega _{01}$ and $\omega _{12}$ correspond to the $\ket{0} \leftrightarrow \ket{1}$ and $\ket{1} \leftrightarrow \ket{2}$ transition frequencies, respectively. As shown in Fig.~\ref{fig:1}(d), here we choose the time-dependent parameters $\Omega_{1}(t)$, $\Omega_{2}(t)$ and $\Delta (t)$ to share a common cosine envelope.

\textit{Appendix B: Qutrit X gate}---%
In the single-photon resonant regime, the general form of the dynamical invariant $I(t)$ for a three-level system can be written as~\cite{chen2012engineering}
\begin{equation}
I(t) = \frac{\hbar \Omega_{0}}{2} 
\begin{pmatrix}
 0 & \cos \gamma \sin \beta  & -i\sin \gamma \\
 \cos \gamma \sin \beta & 0 & \cos \gamma \cos \beta \\
 i\sin \gamma & \cos \gamma \cos \beta  & 0 
\end{pmatrix},
\end{equation}
with the corresponding eigenstates given by
\begin{equation}\label{eqs13}
\begin{aligned}
\ket{\phi_{0}} & = \begin{pmatrix}
\cos \gamma \cos \beta   \\
-i \sin \gamma  \\
-\cos \gamma \sin \beta  
\end{pmatrix},
\\
\ket{\phi_{\pm}} & = \frac{1}{\sqrt{2}} \begin{pmatrix}
\sin \gamma \cos \beta \pm i\sin \beta   \\
-i \cos \gamma  \\
-\sin \gamma \sin \beta  \pm i\cos \beta
\end{pmatrix},
\end{aligned}
\end{equation}
where $\Omega_{0}$ is a frequency scaling constant. The auxiliary parameters $\gamma(t)$ and $\beta(t)$ are related to the drive envelopes $\Omega_{1,2}(t)$ via the consistency relations:
\begin{equation}\label{eqs12}
\begin{aligned}
\dot{\gamma} &= (\Omega_{1} \cos \beta - \Omega_{2} \sin \beta)/2  ,
\\
\dot{\beta} &=  \tan \gamma (\Omega_{1} \sin \beta + \Omega_{2} \cos \beta)/2. 
\end{aligned}
\end{equation}
The LR phases for $ \ket{\phi_{j}(t)} $ can be expressed respectively as $\theta_{0}(T) = 0$, $\theta_{\pm}(T) = \mp \int_{0}^{T}( \dot{\beta}/ \sin \gamma) dt$.

In order to realize the qutrit X gate, we impose the boundary conditions:
\begin{equation}\label{eqs15}
\beta (0) = 0, \quad \beta (T) = \pi / 2, \quad \gamma (0) = 0, \quad \gamma (T) = 0.
\end{equation}
Under these constraints, the time-evolution operator in Eq.~\eqref{eq:inv} simplifies to a function of LR phases $\theta (T) = \theta_{-} (T)$:
\begin{equation}
U(\theta (T)) = \begin{pmatrix}
 0 & i\sin \theta (T)   & \cos \theta (T) \\
 0 &   \cos \theta (T)  & i\sin \theta (T) \\
 -1 & 0 & 0
\end{pmatrix}.
\end{equation}

Setting $\theta (T)= -3\pi/2$ yields $ U ( -3\pi/2 ) = - \ket{2} \bra{0} + i \ket{0} \bra{1} + i\ket{1} \bra{2}$, which corresponds to the $\mathrm{X}^{-1}$ gate up to a diagonal phase gate. Since X and $\mathrm{X}^{-1}$ are time-reversal partners, the control pulses for the X gate are obtained by time-reversing the $\mathrm{X}^{-1}$ sequence ($t \to T-t$). Additionally, setting $\theta (T)= -\pi$ yields $ U ( -\pi ) = - \ket{2} \bra{0} - \ket{1} \bra{1} - \ket{0} \bra{2}$, which is equivalent to the $X_{02}$ gate. To satisfy the boundary conditions, here we adopt the following polynomial ansatz for $\gamma(t)$ and $\beta(t)$:
\begin{equation}
\begin{aligned}
\gamma (t) &= \lambda (\frac{t}{T})^{3} (1-\frac{t}{T})^{3},
\\
\dot{\beta}(t) & = 1386 \frac{\pi}{T}(\frac{t}{T})^{5} (1-\frac{t}{T})^{5},
\end{aligned}
\end{equation}
where $\lambda$ serves as a tunable parameter used to target a specific LR phase. We select $\lambda = 31.5146$ ($\lambda = 48.8597$) to yield an LR phase of $\theta = -3\pi/2$ ($-\pi$), corresponding to the condition for the qutrit X gate ($X_{02}$ gate). The resulting Rabi frequencies $\Omega_{1,2} (t)$ for the X gate, derived via Eqs.~(\ref{eqs12}), are plotted in Fig.~\ref{fig:1}(b).

\textit{Appendix C: Comparison with SU(2) decomposition}---%
To demonstrate the superior compilation efficiency of our direct SU(3) control framework, we compare it against conventional SU(2) decomposition schemes across three critical metrics: (i) arbitrary SU(3) compilation, (ii) Clifford group synthesis, and (iii) H-gate implementation.

(i) Conventional schemes typically employ the Givens decomposition~\cite{lindon2023complete}, where an arbitrary $U \in \mathrm{SU}(3)$ is expressed as a product of three SU(2) rotations $U_{ij}$ acting on distinct two-level subspaces $\{\ket{i}, \ket{j}\}$: $U = U_{12}U_{01}U_{12}$.
Since each $U_{ij}$ requires at most 2 native $\pi/2$ pulses interspersed with virtual phase gates with the U3 decomposition~\cite{morvan2021qutrit}, realizing a general SU(3) operation typically necessitates up to 6 native pulses.

We propose a compact interleaved sequence, given by Eq.~\eqref{Eq:decom}, which effectively reduces the native pulse count by approximately 1/3 compared to conventional decompositions. 
This sequence introduces 10 independent parameters, providing sufficient degrees of freedom to cover the 8-dimensional SU(3) manifold. 
From a Lie algebra perspective, the diagonal phase gates provide the two diagonal generators of $\mathfrak{su}(3)$. Successive commutators of H with $\text{Z}_{\phi_j}$ generate the remaining six off-diagonal elements, spanning the full algebra. 
To verify the universality of our scheme, we performed numerical decomposition on a dataset of 1,000 random SU(3) unitaries, as shown in Fig.~\ref{fig:5}(b).
It is noticed that realizing arbitrary SU(3) gates requires extensive hardware recalibration and resource-intensive characterization techniques, such as quantum process tomography or cross-entropy benchmarking \cite{cho2024direct}.

(ii) High-fidelity Clifford gates are essential for fault-tolerant quantum computing. Conventional SU(2) decompositions require an average of 5.25 native pulses ($\pi/2$ gates) or 3.325 pulses ($\pi/2$ and $\pi$ gates) per Clifford operation \cite{liu2023performing,morvan2021qutrit}.
The standard minimal generating set \{H, S\} is also inefficient, averaging 4.06 H and 4.89 S gates. 
In contrast, the redundant generating set \{H, S, X, Z\} yields an average composition of 1.542 H gates, 1.625 S gates, 0.125 X gates, and 1.648 Z gates, which reduces the native pulse count to 1.667 per Clifford operation. This corresponds to a 68.2\% and 49.9\% reduction in circuit depth compared to the respective SU(2)-based schemes, offering a substantially more efficient and practical alternative.

(iii) Given that the H gate is the most widely used operation in quantum algorithms, its efficient implementation is of critical importance. Table~\ref{tab:compare} contrasts our single-step H gate with recent SU(2)-based realizations. For a fair comparison, all reported fidelities are converted to process fidelities $\mathcal{F}_P$. Our protocol achieves a 35-ns duration, a 3- to 5-fold speedup over conventional methods, while maintaining state-of-the-art fidelity.

\begin{table}[h]
\centering
\caption{Qutrit H gate implementations.}
\label{tab:compare}
\setlength{\tabcolsep}{3.5pt}
\renewcommand{\arraystretch}{1.2}
\begin{tabular}{lcccc}
\hline\hline
 & Method & $\tau$ (ns) & $\mathcal{F}_P$ (\%) & $T_1^{01/12}$, $T_2^{01/12}$ ($\mu$s) \\
\hline
Ref. \cite{morvan2021qutrit}& SU(2) & 124 & 99.4 & 60/36, 60/27 \\
Ref. \cite{roy2023two} & SU(2) & 142 & 97.5 & 48/22, 4.5/2.0 \\
Ref. \cite{liu2023performing} & SU(2) & 190 & 98.4 & 180/101, 76/37 \\
\textbf{Ours} & \textbf{SU(3)} & \textbf{35} & \textbf{99.4} & 61/28, 4.6/4.4 \\
\hline\hline
\end{tabular}
\end{table}
\end{document}


\title{Supplementary Material for 
            ``Efficient Implementation of a Single-Qutrit Gate Set via Coherent Control"
            }

\author{Xiang-Min Yu}
\thanks{These authors contributed equally.}
\affiliation{National Laboratory of Solid State Microstructures, School of Physics, Nanjing University, Nanjing 210093, China}
\affiliation{Shishan Laboratory, Nanjing University, Suzhou 215163, China}
\affiliation{Jiangsu Key Laboratory of Quantum Information Science and Technology, Nanjing University, Suzhou 215163, China}
\affiliation{Synergetic Innovation Center of Quantum Information and Quantum Physics, University of Science and Technology of China, Hefei, Anhui 230026, China}
\affiliation{Hefei National Laboratory, Hefei 230088, China}

\author{Xiang Deng}
\thanks{These authors contributed equally.}
\affiliation{National Laboratory of Solid State Microstructures, School of Physics, Nanjing University, Nanjing 210093, China}
\affiliation{Shishan Laboratory, Nanjing University, Suzhou 215163, China}
\affiliation{Jiangsu Key Laboratory of Quantum Information Science and Technology, Nanjing University, Suzhou 215163, China}

\author{Wen Zheng}
\email{zhengwen@nju.edu.cn}
\affiliation{National Laboratory of Solid State Microstructures, School of Physics, Nanjing University, Nanjing 210093, China}
\affiliation{Shishan Laboratory, Nanjing University, Suzhou 215163, China}
\affiliation{Jiangsu Key Laboratory of Quantum Information Science and Technology, Nanjing University, Suzhou 215163, China}

\author{Wei Xin}
\thanks{These authors contributed equally.}
\affiliation{National Laboratory of Solid State Microstructures, School of Physics, Nanjing University, Nanjing 210093, China}
\affiliation{Shishan Laboratory, Nanjing University, Suzhou 215163, China}
\affiliation{Jiangsu Key Laboratory of Quantum Information Science and Technology, Nanjing University, Suzhou 215163, China}
\author{Tao Zhang}
\affiliation{National Laboratory of Solid State Microstructures, School of Physics, Nanjing University, Nanjing 210093, China}
\affiliation{Shishan Laboratory, Nanjing University, Suzhou 215163, China}
\affiliation{Jiangsu Key Laboratory of Quantum Information Science and Technology, Nanjing University, Suzhou 215163, China}
\author{Hanxin Che}
\affiliation{National Laboratory of Solid State Microstructures, School of Physics, Nanjing University, Nanjing 210093, China}
\affiliation{Shishan Laboratory, Nanjing University, Suzhou 215163, China}
\affiliation{Jiangsu Key Laboratory of Quantum Information Science and Technology, Nanjing University, Suzhou 215163, China}
\author{Kun Zhou}
\affiliation{National Laboratory of Solid State Microstructures, School of Physics, Nanjing University, Nanjing 210093, China}
\affiliation{Shishan Laboratory, Nanjing University, Suzhou 215163, China}
\affiliation{Jiangsu Key Laboratory of Quantum Information Science and Technology, Nanjing University, Suzhou 215163, China}
\affiliation{Synergetic Innovation Center of Quantum Information and Quantum Physics, University of Science and Technology of China, Hefei, Anhui 230026, China}
\affiliation{Hefei National Laboratory, Hefei 230088, China}
\author{Haoyu Zhou}
\affiliation{National Laboratory of Solid State Microstructures, School of Physics, Nanjing University, Nanjing 210093, China}
\affiliation{Shishan Laboratory, Nanjing University, Suzhou 215163, China}
\affiliation{Jiangsu Key Laboratory of Quantum Information Science and Technology, Nanjing University, Suzhou 215163, China}
\author{Yangyang Ge}
\affiliation{National Laboratory of Solid State Microstructures, School of Physics, Nanjing University, Nanjing 210093, China}
\affiliation{Shishan Laboratory, Nanjing University, Suzhou 215163, China}
\affiliation{Jiangsu Key Laboratory of Quantum Information Science and Technology, Nanjing University, Suzhou 215163, China}
\author{Zhenchuan Zhang}
\affiliation{Shishan Laboratory, Nanjing University, Suzhou 215163, China}
\affiliation{Jiangsu Key Laboratory of Quantum Information Science and Technology, Nanjing University, Suzhou 215163, China}
\author{Wanli Huang}
\affiliation{Shishan Laboratory, Nanjing University, Suzhou 215163, China}
\affiliation{Jiangsu Key Laboratory of Quantum Information Science and Technology, Nanjing University, Suzhou 215163, China}
\author{Haoyang Cai}
\affiliation{National Laboratory of Solid State Microstructures, School of Physics, Nanjing University, Nanjing 210093, China}
\affiliation{Shishan Laboratory, Nanjing University, Suzhou 215163, China}
\affiliation{Jiangsu Key Laboratory of Quantum Information Science and Technology, Nanjing University, Suzhou 215163, China}
\author{Xianke Li}
\affiliation{National Laboratory of Solid State Microstructures, School of Physics, Nanjing University, Nanjing 210093, China}
\affiliation{Shishan Laboratory, Nanjing University, Suzhou 215163, China}
\affiliation{Jiangsu Key Laboratory of Quantum Information Science and Technology, Nanjing University, Suzhou 215163, China}

\author{Jie Zhao}
\affiliation{National Laboratory of Solid State Microstructures, School of Physics, Nanjing University, Nanjing 210093, China}
\affiliation{Shishan Laboratory, Nanjing University, Suzhou 215163, China}
\affiliation{Jiangsu Key Laboratory of Quantum Information Science and Technology, Nanjing University, Suzhou 215163, China}
\author{Xinsheng Tan}
\affiliation{National Laboratory of Solid State Microstructures, School of Physics, Nanjing University, Nanjing 210093, China}
\affiliation{Shishan Laboratory, Nanjing University, Suzhou 215163, China}
\affiliation{Jiangsu Key Laboratory of Quantum Information Science and Technology, Nanjing University, Suzhou 215163, China}
\affiliation{Synergetic Innovation Center of Quantum Information and Quantum Physics, University of Science and Technology of China, Hefei, Anhui 230026, China}
\affiliation{Hefei National Laboratory, Hefei 230088, China}

\author{Yu Zhang}
\affiliation{National Laboratory of Solid State Microstructures, School of Physics, Nanjing University, Nanjing 210093, China}
\affiliation{Shishan Laboratory, Nanjing University, Suzhou 215163, China}
\affiliation{Jiangsu Key Laboratory of Quantum Information Science and Technology, Nanjing University, Suzhou 215163, China}
\author{Shao-Xiong Li}
\email{shaoxiong.li@nju.edu.cn}
\affiliation{National Laboratory of Solid State Microstructures, School of Physics, Nanjing University, Nanjing 210093, China}
\affiliation{Shishan Laboratory, Nanjing University, Suzhou 215163, China}
\affiliation{Jiangsu Key Laboratory of Quantum Information Science and Technology, Nanjing University, Suzhou 215163, China}
\affiliation{Synergetic Innovation Center of Quantum Information and Quantum Physics, University of Science and Technology of China, Hefei, Anhui 230026, China}
\affiliation{Hefei National Laboratory, Hefei 230088, China}
\author{Yang Yu}
\email{yuyang@nju.edu.cn}
\affiliation{National Laboratory of Solid State Microstructures, School of Physics, Nanjing University, Nanjing 210093, China}
\affiliation{Shishan Laboratory, Nanjing University, Suzhou 215163, China}
\affiliation{Jiangsu Key Laboratory of Quantum Information Science and Technology, Nanjing University, Suzhou 215163, China}
\affiliation{Synergetic Innovation Center of Quantum Information and Quantum Physics, University of Science and Technology of China, Hefei, Anhui 230026, China}
\affiliation{Hefei National Laboratory, Hefei 230088, China}

\date{\today}
\maketitle
\tableofcontents

~\\
~\\

This Supplemental Material provides comprehensive theoretical derivations and detailed experimental procedures supporting the main text. For completeness, specific implementation details and comparative advantages of the proposed scheme—briefly discussed in the End Matter—are revisited here with expanded numerical simulations and greater technical depth.
The first section details the construction of a Clifford-based single-qutrit gate set through coherent SU(3) dynamics control, followed by an extension of efficient Ramsey interferometry and parity determination in quantum circuits in the second section. The third section presents additional experimental sample information, demonstrating the broad applicability of our approach without loss of generality. The fourth section elaborates on the experimental details, highlighting the ease with which our proposed scheme can be adapted to other quantum platforms. Our coherent control scheme paves the way for enhancing performance in qutrit processors, providing exciting prospects for the advancement of high-dimensional quantum hardware and algorithms.

\section{Theoretical derivations for qutrit gates } \label{sec:1}
Consider a qutrit modeled as a cascaded three-level system with basis states $\{ \ket{0}, \ket{1}, \ket{2} \}$. When the qutrit is simultaneously driven by two coherent pulses under two-photon resonance regime, the Hamiltonian takes the form
\begin{equation} \label{eqs1}
H/\hbar = \Delta \ket{1} \bra{1} +  \frac{1}{2} (\Omega_{1} \ket{0} \bra{1} + \Omega_{2} \ket{1} \bra{2} + h.c.),
\end{equation} 
where $\Delta$ denotes the single-photon detuning, and $\Omega_{1}$, $ \Omega_{2} $ are the Rabi frequencies of coherent drives for the transitions $\ket{0} \leftrightarrow \ket{1} $ and $\ket{1} \leftrightarrow \ket{2} $, respectively. Here, we present a method for  identifying appropriate control pulses $\Delta$, $\Omega_{1}$, and $\Omega_{2}$ necessary for generating qutrit H gate and X gate. Notably, it proves convenient to achieve this goal by directly addressing SU(3) dynamics.

\subsection{Qutrit H gate}
When the single-photon detuning $\Delta$ and Rabi frequencies $\Omega_{1}$, $\Omega_{2}$ remain constant over time, i.e., the Hamiltonian Eq. \eqref{eqs1} is time-independent, the corresponding time-evolution operator $U(T)$ can be formulated as the exponential map of the Hamiltonian \cite{zhang2022robust, zhang2024high}, that is 
\begin{equation} \label{eqs2}
U(T) = \begin{pmatrix}
 U_{00} &  U_{01} & U_{02} \\
 U_{10} &  U_{11} & U_{12} \\
 U_{20} &  U_{21} & U_{22}
\end{pmatrix} ,
\end{equation} 
with
\begin{equation} \label{eqs3}
\begin{aligned}
U_{00} &= \cos^{2} \theta + e^{-i \delta} ( \cos \frac{A}{2} + i \frac{\Delta}{\Omega} \sin \frac{A}{2} ) \sin^{2} \theta \\
U_{11} &= e^{-i \delta} (\cos \frac{A}{2} - i \frac{\Delta}{\Omega} \sin \frac{A}{2} )  \\
U_{22} &= \sin^{2} \theta + e^{-i \delta} ( \cos \frac{A}{2} + i \frac{\Delta}{\Omega} \sin \frac{A}{2} ) \cos^{2} \theta  \\
U_{01} &= U_{10} = -i e^{-i \delta} \frac{\Omega_{0}}{\Omega} \sin \theta \sin \frac{A}{2}  \\
U_{12} &= U_{21} = -i e^{-i \delta} \frac{\Omega_{0}}{\Omega} \cos \theta \sin \frac{A}{2}  \\
U_{02} &= U_{20} = e^{-i \delta} ( \cos \frac{A}{2} + i \frac{\Delta}{\Omega} \sin \frac{A}{2} - e^{i \delta} ) \sin \theta \cos \theta
\end{aligned}
\end{equation} 
where the reduced parameters $\theta$, $\Omega_{0}$, $\Omega$, $A$ and $\delta$ are defined as $\tan \theta = \Omega_{1}/\Omega_{2}$, $\Omega_{0} = \sqrt{\Omega_{1}^{2} + \Omega_{2}^{2}}$, $\Omega = \sqrt{\Omega_{0}^{2} + \Delta^{2}}$, $A = \Omega T$ and $\delta = \Delta T/2$, respectively. It is worth noting that any $3 \times 3$ unitary operator with equal-modulus matrix elements, i.e., $|U_{ij}| = 1 / \sqrt{3}$, can be represented as either a qutrit H gate or its inverse, sandwiched between two diagonal phase gates \cite{shlyakhov2018quantum}. Consequently, in order to determine the parameters $(\Omega_{1} T, \Omega_{2} T, \Delta T)$ that generate the qutrit H gate, we have to solve the equation $|U_{ij}| = 1 / \sqrt{3}$. It is straightforward to observe that $\Omega_{1} = \Omega_{2}$, i.e., $\tan \theta = 1$ , by enforcing the condition $|U_{01}| = |U_{12}|$. Under this constraint, it remains only to identify the parameters $(A, \delta)$, the conditions are consequently reduced to
\begin{subequations}\label{eqs4}
\begin{align}
\delta^{2} &= (A/2)^{2}( 1-\frac{2}{3 \sin^{2}(A/2)} ), \\
\cos (A/2) &\cos \delta + (2\delta /A) \sin (A/2) \sin \delta = 0.
\end{align}
\end{subequations}
A range of solutions can be identified from the intersection points of the geometric representations associated with Eqs.~(\ref{eqs4}). Among these solutions, we select the one with the smallest value of $A$, as it yields the minimal Rabi frequencies for a given evolution time $T$, that is $|A| = 4.0410$ and $|\delta| = 0.8525$, i.e. $\Omega_{1}T = \Omega_{2}T = 2.5906$ and $|\Delta T| = 1.7050$. For a H gate with evolution time $T = 35$ ns, $\Delta$ should take a positive value, we thus deduce the parameters $\Omega_{1}/2\pi = \Omega_{2}/2\pi = 11.7801$ MHz and $\Delta/2\pi = 7.7531$ MHz. At this point, the qutrit H gate can be explicitly expressed as
\begin{equation}  \label{eqs5}
\mathrm{H} = \begin{pmatrix}
 1 & 0 & 0 \\
 0 & e^{i(\frac{2\pi}{3} + \delta  )}  & 0 \\
 0 & 0 & e^{-i\frac{2\pi}{3}}
\end{pmatrix}
\cdot U(T) \cdot
\begin{pmatrix}
 e^{-i\frac{\pi}{6}} & 0 & 0 \\
 0 & e^{i(\frac{\pi}{2} + \delta  )}  & 0 \\
 0 & 0 & e^{-i\frac{5\pi}{6}}
\end{pmatrix}.
\end{equation} 

The above derivation assumes a time-independent Hamiltonian, with $\Omega_1$ and $\Omega_2$ modeled as square pulses. However, as discussed in the main text, it is necessary to require the drive amplitudes to start and end at zero in practice.
In addition, square pulses exhibit broad spectral bandwidths \cite{Martinis2014}, making them more susceptible to waveform distortion and more likely to induce significant cross-coupling.
Therefore, such pulses are not suitable for the present experimental setup. Note that if $\Delta$ and $\Omega_{1}$, $\Omega_{2}$ share the same time dependence, i.e., the qutrit is driven by phase-modulated chirp pulses, it is also feasible to implement qutrit H gate as soon as the following conditions hold, i.e. $\int_{0}^{T}\Omega_{1} (t)dt = \int_{0}^{T}\Omega_{2} (t)dt = 2.5906 $ and $\int_{0}^{T} \Delta(t) dt = 1.7050$. In this paper, the chirped driving pulses applied in the laboratory frame take the following form
\begin{subequations}\label{eqs6}
\begin{align}
\tilde{\Omega}_{1}(t) &=  \Omega_{1}(t) \cos\left(\omega _{01}t  + \int_{0}^{t} \Delta (\tau)d\tau\right),  \\
\tilde{\Omega}_{2}(t) &=  \Omega_{2}(t) \cos\left(\omega _{01}t  - \int_{0}^{t} \Delta (\tau)d\tau\right). 
\end{align}
\end{subequations}
where $\omega _{01}$ and $\omega _{12}$ correspond to the $\ket{0} \leftrightarrow \ket{1}$ and $\ket{1} \leftrightarrow \ket{2}$ transition frequencies, respectively. The time-dependent parameters $\Omega_{1}(t)$, $\Omega_{2}(t)$ and $\Delta (t)$ share a common envelope, characterized by 5-ns Gaussian rising and falling edges with a flat-top profile. The corresponding population dynamics, obtained via numerical simulation using QuTiP \cite{johansson2012qutip,lambert2024}. Note that by introducing a parity-like operator $P = \ket{0} \bra{2} + \ket{2}\bra{0}$, the corresponding Hamiltonian $H$ exhibits $\mathcal{P}$-symmetry, i.e. $PH(t)P = H(t)$, as illustrated in Fig.~\ref{fig:s1}.

\renewcommand{\thefigure}{S\arabic{figure}} 
\begin{figure}[ht]
    \centering
    \includegraphics[width = 0.900\textwidth, height = 0.175\textheight]{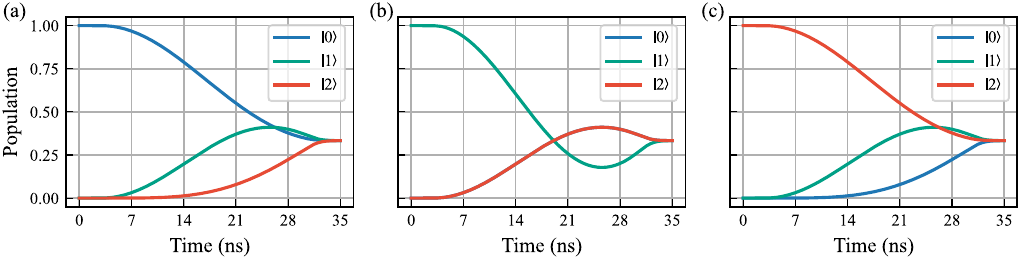}
    \caption{\justifying Qutrit state population dynamics during the implementation of the $H$ gate, with the qutrit initialized in (a) $\ket{0}$, (b) $\ket{1}$, and (c) $\ket{2}$, respectively. The population dynamics of states $\ket{0}$ and $\ket{2}$ are mirror images of each other.}
    \label{fig:s1}
\end{figure}

It is worth noting that the conjugate term of the H gate can be natively generated by taking $\Delta$ with a negative value. Then, the $\mathrm{H}^{-1}$ gate can be obtained with two phase gates, stated as
\begin{equation}  \label{eqs7}
\mathrm{H}^{-1} = \begin{pmatrix}
 1 & 0 & 0 \\
 0 & e^{i(\frac{\pi}{3} + \delta  )}  & 0 \\
 0 & 0 & e^{i\frac{2\pi}{3}}
\end{pmatrix}
\cdot U(T) \cdot
\begin{pmatrix}
 e^{i\frac{\pi}{6}} & 0 & 0 \\
 0 & e^{i(\frac{\pi}{2} + \delta  )}  & 0 \\
 0 & 0 & e^{i\frac{5\pi}{6}}
\end{pmatrix}
\end{equation} 
This unique operation enables high-performance quantum algorithms such as Ramsey interferometry and parity determination, enhancing its efficiency in practice, particularly in reducing quantum circuit compiler overheads. 
In addition, by setting $\Delta = 0$ and $A = 2\pi$, we can realize arbitrary operations within the $\{ \ket{0}, \ket{2} \}$ subspace, even though the corresponding transition is nominally forbidden.

\subsection{Qutrit X gate}
As stated in the main text, the unitary evolution governed by a time-independent real Hamiltonian preserves time-reversal symmetry. Consequently, it is not feasible to employ the time-independent Hamiltonian to implement the qutrit X gate, which intrinsically breaks time-reversal symmetry. Here, we consider the case where the qutrit is resonantly driven by two time-varying pulses. The corresponding Hamiltonian can be simplified as
\begin{equation} \label{eqs8}
H/\hbar = \frac{1}{2} ( \Omega_{1}(t)\left | 0 \right \rangle  \left \langle 1 \right |  
+ \Omega_{2}(t) \left | 1 \right \rangle  \left \langle 2 \right | )  + h.c. .
\end{equation} 
Since the time-evolution operator can only be given in terms of the Dyson series according to the time-dependent Schr{\"o}dinger equation, it is challenging to determine the precise form of the Rabi frequencies directly to obtain the desired qutrit X gate owing to the complexity of the commutation relations. In particular, Lewis and Riesenfeld have proposed a dynamical invariant theory to address time-dependent Schr{\"o}dinger equation, thereby enabling the derivation of the time-evolution operator \cite{lewis1969,chen2011lewis}. In the context of Lewis-Riesenfeld (LR) invariant theory, if a quantum system evolving with a time-dependent Hamiltonian $H(t)$ possesses a dynamical invariant $I(t)$ that satisﬁes $ \partial I(t) / \partial t + i/\hbar [ H(t), I(t) ] = 0 $, then the associated time-evolution operator can be expressed as 
\begin{equation}
U(T) = \sum_{j}e^{i\theta_{j}(T)} \ket{\phi_{j}(T)} \bra{\phi_{j}(0)}  ,
\end{equation}
where $ \ket{\phi_{j}(t)} $ represent the eigenstates of the dynamical invariant $I(t)$, satisfying $I(t) \ket{\phi_{j}(t)} = \lambda _{j} \ket{\phi_{j}(t)}$, and $j=0,1,2$. Note that $\lambda _{j}$ are time-independent constants. The accumulated LR phases $\theta_{j}(T)$ are deﬁned as
\begin{equation}
\theta_{j}(T) = \frac{1}{\hbar} \int_{0}^{T}
\bra{\phi_{j}(t)} i\hbar \frac{\partial}{\partial t}- H(t) \ket{\phi_{j}(t)} dt .
\end{equation}

In particular, the general form of the dynamical invariant $I(t)$ for a three-level system, as characterized by Hamiltonian (\ref{eqs8}), has already been established \cite{chen2012engineering}, which can be expressed as
\begin{equation}
I(t) = \frac{\hbar}{2}  \Omega_{0} \begin{pmatrix}
 0 & \cos \gamma \sin \beta  & -i\sin \gamma \\
 \cos \gamma \sin \beta & 0 & \cos \gamma \cos \beta \\
 i\sin \gamma & \cos \gamma \cos \beta  & 0 
\end{pmatrix} ,
\end{equation}
where $\Omega_{0}$ denotes an arbitrary constant with units of frequency to maintain the dynamical invariant $I(t)$ with dimensions of energy. The time-dependent auxiliary parameters $\gamma(t)$ and $\beta(t)$ are subject to specific constraints, viz.,
\begin{subequations}\label{eqs12}
\begin{align}
\dot{\gamma} &= (\Omega_{1} \cos \beta - \Omega_{2} \sin \beta)/2  ,
\\
\dot{\beta} &=  \tan \gamma (\Omega_{1} \sin \beta + \Omega_{2} \cos \beta)/2  , 
\end{align}
\end{subequations}
which can in turn be employed to design Rabi frequencies $\Omega_{1}(t)$ and $\Omega_{2}(t)$.
The eigenstates associated with the dynamical invariant $I(t)$ are cast as 
\begin{subequations}\label{eqs13}
\begin{align}
\ket{\phi_{0}} & = \begin{pmatrix}
\cos \gamma \cos \beta   \\
-i \sin \gamma  \\
-\cos \gamma \sin \beta  
\end{pmatrix},
\\
\ket{\phi_{1,2}} & = \begin{pmatrix}
\sin \gamma \cos \beta \pm i\sin \beta   \\
-i \cos \gamma  \\
-\sin \gamma \sin \beta  \pm i\cos \beta
\end{pmatrix},
\end{align}
\end{subequations}
which correspond to the eigenvalues $\lambda_{0}=0$ and $\lambda_{\pm }=\pm \hbar \Omega _{0}/2$. In this case, the related LR phases for $ \ket{\phi_{j}(t)} $ can be written respectively as
\begin{subequations}\label{eqs14}
\begin{align}
\theta_{0}(T) &= 0 ,
\\
\theta_{1,2}(T) &= \mp \int_{0}^{T}( \dot{\beta}/ \sin \gamma) dt. 
\end{align}
\end{subequations}

Note that the time-evolution operator for a three-level system is completely determined by the accumulated LR phases $\theta_{\pm}$, once appropriate boundary conditions for the time-dependent auxiliary parameters for $\gamma(t)$ and $\beta(t)$ are given. Consequently, a specific qutrit gate operation can be effectively implemented with a further tailored design of the $\gamma(t)$ and $\beta(t)$ to regulate the LR phases $\theta_{\pm}$. 

In order to realize the qutrit X gate, we consider the following boundary conditions:
\begin{subequations}\label{eqs15}
\begin{align}
\beta (0) & = 0, \quad \beta (T) = \pi / 2 ,
\\
\gamma (0) & = 0, \quad \gamma (T) = 0 .
\end{align}
\end{subequations}
with $T$ being the operation time. At this point, the time-evolution operator can be expressed as a function of LR phases $\theta (T) = \theta_{-} (T)$, viz.,
\begin{equation}
U(\theta (T)) = \begin{pmatrix}
 0 & i\sin \theta (T)   & \cos \theta (T) \\
 0 &   \cos \theta (T)  & i\sin \theta (T) \\
 -1 & 0 & 0
\end{pmatrix}.
\end{equation}

It is obvious that when $\theta (T)= -\pi$, the time-evolution operator is given by $ U ( -\pi ) = - \ket{2} \bra{0} - \ket{1} \bra{1} - \ket{0} \bra{2}$, which is equivalent to  $X_{02}$, i.e. $0 \leftrightarrow 2$ SWAP gate with the global $\pi$-phase omitted. Furthermore, when $\theta (T)= -3\pi/2$, the time-evolution operator $ U ( -3\pi/2 ) = - \ket{2} \bra{0} + i \ket{0} \bra{1} + i\ket{1} \bra{2}$ indicates a non-reciprocal chiral population transportation $ 0\to 2\to 1 \to 0$, which differs from $\mathrm{X}^{-1}$ by only a diagonal phase gate. The phase gate can be addressed via a virtual phase operation, as will be detailed in Sec.~\ref{sec:s3_vz}. Since the X gate and $\mathrm{X}^{-1}$ gate are time-reversal counterparts, the driving pulses for the X gate can be obtained by applying time inversion, $t \to T-t$, to the $\mathrm{X}^{-1}$ gate pulses about the midpoint $T/2$ of the operation period. As a result, it is feasible to realize a qutrit X gate based on LR invariant theory, provided that the appropriate boundary conditions are set in place.

\renewcommand{\thefigure}{S\arabic{figure}} 
\begin{figure}[ht]
    \centering
    \includegraphics[width = 0.900\textwidth, height = 0.176\textheight]{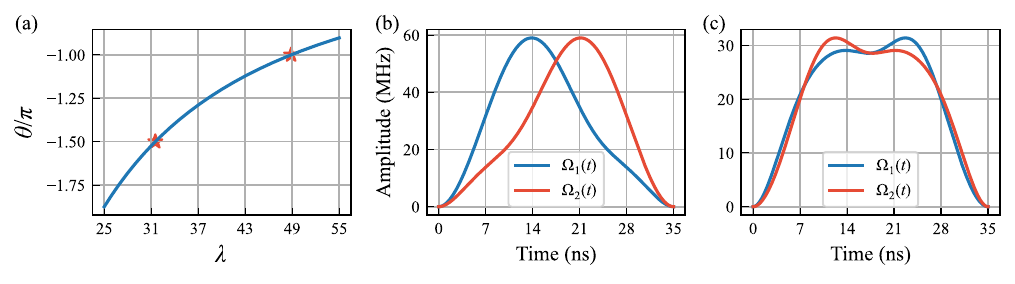}
    \caption{\justifying Qutrit X-type gate pulses design via LR invariant. (a) The LR phase $\theta$ as a function of the tunable parameter $\lambda$. The two points marked with stars correspond to LR phases of -1.5$\pi$ and -$\pi$, which yield the X gate and $\mathrm{X}_{02}$ gate , respectively, with the corresponding values $\lambda = 31.5146 $ and $\lambda = 48.8597 $. (b), (c) The Rabi frequencies $\Omega_{1}$ and $\Omega_{2}$ designed for the X gate and $\mathrm{X}_{02}$ gate, respectively, in terms of the corresponding $\lambda$ values. Note that $\Omega_{1}$ and $\Omega_{2}$ are time-reversed counterparts of each other.}
    \label{fig:s2}
\end{figure}

Prior to specifying the particular forms of the time-dependent auxiliary parameters $\gamma(t)$ and $\beta(t)$ for the qutrit X gate, it is necessary to take additional boundary conditions into account beyond those outlined in Eqs.~(\ref{eqs15}), particularly in light of the experimental constraints. In experiments, we are concerned with situations where the driving pulses $\Omega_{1,2} (t)$ and their corresponding first-order derivatives $\dot{\Omega}_{1,2} (t)$  are set to zero at both the initial and final times. Accordingly, we impose the following boundary conditions:
\begin{subequations} \label{eqs18}
\begin{align}
\Omega _{1}( 0 )  =  \Omega _{1}( T )  = 0 ,
\\ 
\Omega _{2} ( 0 )  =  \Omega _{2} ( T )  = 0,
\\
\dot{\Omega}_{1} ( 0 )  =  \dot{\Omega}_{1} ( T )  = 0 ,
\\
\dot{\Omega}_{2} ( 0 )  =  \dot{\Omega}_{2} ( T )  = 0 .
\end{align}
\end{subequations}
Note that extra caution is required at the final time, since the boundary conditions suggest a potential divergence of the driving pulses $\Omega_{1,2} (t)$ due to $ \cot \gamma \longmapsto \infty (\gamma \rightarrow 0) $.
To address this challenge, it is essential to ensure that the product $\beta(t) \cot \gamma(t)$ remains finite. In this paper, we simply adopt a polynomial ansatz to interpolate the functions $\gamma(t)$ and $\beta(t)$ at intermediate times, i.e., $ \gamma  ( t ) = {\textstyle \sum_{j}}a_{j}t^{j} $ and $ \beta ( t ) = {\textstyle \sum_{j}}b_{j}t^{j} $, in which the coefficients are determined in terms of the aforementioned boundary.
Thus, a set of polynomials $\gamma(t)$ and $\beta(t)$ can be formulated in a proposal form denoted as
\begin{subequations}
\begin{align}
\gamma (t) = & \lambda (\frac{t}{T})^{3} (1-\frac{t}{T})^{3}  
\\
\beta (t) = & 231\pi (\frac{t}{T})^{6}- 990 \pi (\frac{t}{T})^{7} + \frac{3465 \pi}{2} \pi (\frac{t}{T})^{8} - 1540 \pi (\frac{t}{T})^{9} \\
& +  693 \pi (\frac{t}{T})^{10} - 126 \pi (\frac{t}{T})^{11}
\end{align}
\end{subequations}
where $\lambda$ acts as an adjustable time-independent auxiliary parameter that is utilized to identify the specific LR phase concerned in our proposal, as shown in Fig.~\ref{fig:s2}(a).
We have selected the auxiliary parameter $\lambda = 31.5146$ ($\lambda = 48.8597$) to ensure that the LR phase $\theta$ is calibrated to $-3\pi/2$ ($-\pi$), which is precisely the condition required to implement the qutrit X gate ($X_{02}$ gate). The corresponding Rabi frequencies $\Omega_{1,2} (t)$ for X gate and $X_{02}$ gate,
which are derived from Eqs.~(\ref{eqs12}), are depicted in Fig.~\ref{fig:s2}(b) and Fig.~\ref{fig:s2}(c), respectively. Furthermore, the function $\beta(t)$ is chosen such that $\dot{\beta}(t)  = (1386 \frac{\pi}{T})[\frac{t}{T}(1-\frac{t}{T})]^{5}$, ensuring that Eqs.~(\ref{eqs18}) holds. Noting the symmetry properties $\gamma(T-t)=\gamma(t)$, $\dot{\gamma} (T-t) =- \dot{\gamma} (t)$, $\beta (T-t) = \pi/2 - \beta  (t)$, and
$\dot{\beta} (T-t) = \dot{\beta } (t)$,  it follows that $\Omega _{1}(T-t)= \Omega _{2}(t),~ \Omega _{2}(T-t)=\Omega _{1}(t)$. Similarly, by introducing the parity-like operator $P = \ket{0} \bra{2} + \ket{2}\bra{0}$,  the corresponding Hamiltonian $H$ satisfies $\mathcal{PT}$-symmetry, i.e., $PH(t)P=H(T-t)$. Fig.~\ref{fig:s3} present the population evolution during the application of these gates when the qutrit is separately initialized in states $\ket{0}$, $\ket{1}$ and $\ket{2}$. The results of our numerical simulations are in excellent agreement with the LR theory.

\renewcommand{\thefigure}{S\arabic{figure}} 
\begin{figure}[ht]
    \centering
    \includegraphics[width = 0.900\textwidth, height = 0.340\textheight]{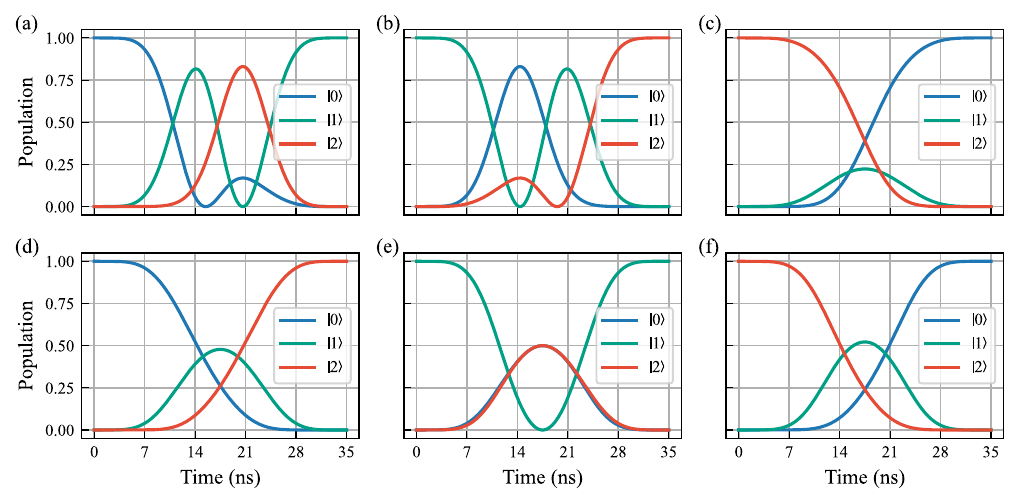}
    \caption{\justifying Qutrit state population dynamics during the X-type gate,  demonstrating the characteristic features of $\mathcal{PT}$-symmetry. (a)–(c) show the dynamics for the X gate, and (d)–(f) for the $\mathrm{X}_{02}$ gate, with the qutrit initialized in $\ket{0}$, $\ket{1}$, and $\ket{2}$, respectively.}
    \label{fig:s3}
\end{figure}

In practice, deviations between experimental and theoretical parameters are inevitable. Therefore, robustness against parameter fluctuations is generally considered a key criterion  for assessing the flexibility of a protocol.
As shown in Fig.~\ref{fig:s3}, the state populations remain nearly unchanged at the initial and final time, suggesting that the invariant-based qutrit X gate is intrinsically robust to control errors, a favorable feature of inverse engineering using LR invariants \cite{chen2012engineering,yu2025quantum}. To further evaluate this robustness, we numerically simulate the gate performance under systematic errors in both the amplitude and frequency of the control pulses. Fig.~\ref{fig:s4}(a) illustrates the variation of gate fidelity with respect to amplitude errors $\eta_{1,2} \in [-0.15, 0.15]$, introduced by modifying the Rabi frequencies as $\Omega_{1,2} \to (1 + \eta_{1,2})\Omega_{1,2}$. Fig.~\ref{fig:s4}(b) explores the impact of detuning errors on the gate fidelity, implemented by shifting the drive frequencies as $\omega_{1,2} \to \omega_{1,2} + \delta_{1,2}$, where the detuning error is defined as $ \delta_{1,2} = 2\pi \zeta_{1,2}/T$ with $\zeta_{1,2} \in [-0.15, 0.15]$ and $T$ being the gate operation time. As demonstrated in Fig.~\ref{fig:s4}, the invariant-based X gate retains high fidelity over a broad error range, highlighting its enhanced robustness.

\renewcommand{\thefigure}{S\arabic{figure}} 
\begin{figure}[ht]
    \centering
    \includegraphics[width = 0.900\textwidth, height = 0.264\textheight]{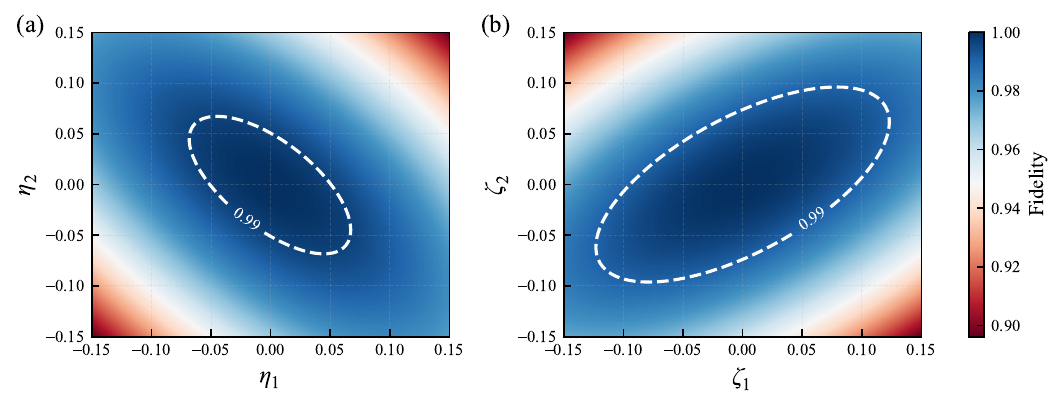}
    \caption{\justifying Robustness of qutrit X gate fidelity against errors in pulse amplitudes and detunings. The gate fidelity as a function of (a) pulse amplitude errors and (b) pulse detuning errors is obtained through numerical simulation.}
    \label{fig:s4}
\end{figure}

It is worth noting that the qutrit X gate corresponds to a three-cycle permutation operation $0\to 1\to 2\to 0$ within the computational subspace. It is well known that there are $3!$ elements in three-element symmetric group $S_{3}$, which correspond to six distinct X-type gates, that is
\begin{equation} \label{eqs17}
\begin{aligned}
\mathrm{I} &= \begin{pmatrix}1 & 0 & 0 \\ 0 & 1 & 0 \\ 0 & 0 & 1 \end{pmatrix},  \quad
\mathrm{X}_{01} = \begin{pmatrix}0 & 1 & 0 \\ 1 & 0 & 0 \\ 0 & 0 & 1 \end{pmatrix}, 
 \quad
\mathrm{X}_{12} = \begin{pmatrix}1 & 0 & 0 \\ 0 & 0 & 1 \\ 0 & 1 & 0 \end{pmatrix}, \\
\mathrm{X}_{02} &= \begin{pmatrix}0 & 0 & 1 \\ 0 & 1 & 0 \\ 1 & 0 & 0 \end{pmatrix},
\quad
\mathrm{X}_{012} = \begin{pmatrix}0 & 0 & 1 \\ 1 & 0 & 0 \\ 0 & 1 & 0 \end{pmatrix},
\quad
\mathrm{X}_{210} = \begin{pmatrix}0 & 
1 & 0 \\ 0 & 0 & 1 \\ 1 & 0 & 0 \end{pmatrix},
\end{aligned}
\end{equation} 
in which I gate, $\mathrm{X}_{012}$(X) gate, $\mathrm{X}_{210}$($\mathrm{X}^{-1}$) gate are even permutations, while $\mathrm{X}_{01}$ gate, $\mathrm{X}_{12}$ gate and $\mathrm{X}_{02}$ gate are odd permutations.
X-type gates play an important role in qutrit-based quantum computing, such as applications in initial state preparation, fast reset, etc. In our theoretical framework, $\mathrm{X}_{012}$ gate, $\mathrm{X}_{210}$ gate and $\mathrm{X}_{02}$ gate have been realized using LR invariant. The $\mathrm{X}_{01}$ gate and $\mathrm{X}_{12}$ gate can be directly implemented via standard SU(2) operations or invariant-based protocol within their respective two-level subspaces. Consequently, any qutrit X-type gate can be implemented in one step, with its parity subsequently verified using a quantum algorithm, as demonstrated in the main text.

\subsection{Virtual Z gates} \label{sec:s3_vz}
Similar to qubit case, a general qutrit diagonal phase gate $\mathrm{Z}_{\phi}$, defined by
\begin{equation}
\mathrm{Z}_{\phi} = \begin{pmatrix}
 1 & 0 & 0\\
 0 & e^{i \phi_{1} }  &0 \\
 0 & 0 & e^{i ( \phi_{1} + \phi_{2}) } 
\end{pmatrix},
\end{equation}
can be implemented virtually by adjusting the phases of the subsequent control pulses. This technique has been demonstrated in recent experiments \cite{yurtalan2020implementation,morvan2021qutrit}.
Such virtual phase gates are realized efficiently at the software level with effectively zero duration, thereby significantly reducing the overall circuit depth \cite{mckay2017efficient}. In the following, we present a concise theoretical analysis for extending this technique in qutrit gates.

We investigate the relationship between two Hamiltonians, denoted as $\tilde{H}$ and $H$, written as
\begin{equation}
\tilde{H} = \begin{pmatrix}
 0 & \frac{\Omega _{1} }{2}e^{i\phi _{1} }   & 0\\
\frac{\Omega _{1} }{2}e^{-i\phi _{1} }  & \Delta  & \frac{\Omega _{2} }{2}e^{i\phi _{2} }\\
 0 & \frac{\Omega _{2} }{2}e^{-i\phi _{2} } & \delta 
\end{pmatrix}, \quad
H = \begin{pmatrix}
0 & \frac{\Omega _{1} }{2} & 0\\
\frac{\Omega _{1} }{2} & \Delta  &  \frac{\Omega _{2} }{2}\\
0 &  \frac{\Omega _{2} }{2} & \delta 
\end{pmatrix},
\end{equation}
which differ only by relative phase shifts in their respective driving pulses.
These Hamiltonians are related via a unitary transformation, expressed as
\begin{equation}
\tilde{H} = \mathrm{Z}^{\dagger}_{\phi} H \mathrm{Z}_{\phi}. 
\end{equation}
Consequently, their corresponding time-evolution operators satisfy
\begin{equation}
\tilde{U} = \mathrm{Z}^{\dagger}_{\phi} U \mathrm{Z}_{\phi}.
\end{equation}
This relation implies that phase shifts in driving pulses effectively conjugate the evolution by a diagonal phase gate, allowing such gates to be virtually commuted through the circuit.
As a result, phase gates can be effectively pushed to the end of the quantum circuit, i.e.
\begin{equation}
\begin{aligned}
U_{1} \mathrm{Z}_{\phi 1} U_{2} \mathrm{Z}_{\phi 2} U_{3} &= 
U_{1} (\mathrm{Z}_{\phi 1} U_{2} \mathrm{Z}^{\dagger}_{\phi 1}) ((\mathrm{Z}_{\phi 1} \mathrm{Z}_{\phi 2}) U_{3} (\mathrm{Z}_{\phi 1} \mathrm{Z}_{\phi 2})^{\dagger}) (\mathrm{Z}_{\phi 1} \mathrm{Z}_{\phi 2})
\\
&= U_{1} \tilde{U}_{2} \tilde{\tilde{U}}_{3} (\mathrm{Z}_{\phi 1} \mathrm{Z}_{\phi 2}),
\end{aligned}
\end{equation}
where $\tilde{U}_{2}= \mathrm{Z}_{\phi 1} U_{2} \mathrm{Z}^{\dagger}_{\phi 1}$ and $\tilde{\tilde{U}}_{3}=(\mathrm{Z}_{\phi 1} \mathrm{Z}_{\phi 2}) U_{3} (\mathrm{Z}_{\phi 1} \mathrm{Z}_{\phi 2})^{\dagger}$
can be implemented by shifting the reference phase of all subsequent single-qutrit operations within a given gate sequence.
Note that, in this case, it is not the bare $U_{2}$ or $U_{3}$ gates that are executed, but rather their phase-shifted counterparts $\tilde{U}{2}$ or $\tilde{\tilde{U}}{3}$.
This compilation strategy therefore requires the hardware to be calibrated to natively support gates with different phase references.
Moreover, the residual phase gates appearing at the end of a circuit can be neglected experimentally, since final measurements typically project onto the computational basis, erasing any remaining phase information.
Hence, with this compilation strategy, phase gates are realized entirely in software, eliminating the need for additional physical pulses.


\subsection{Qutrit Clifford gates}\label{sec:C3gates}
The Clifford group consists of unitary operators that normalize the Pauli group, that is, a unitary gate $C$ belongs to the Clifford group if for any Pauli operator $P$, the conjugated operator $CPC^\dagger$ is also a Pauli operator, up to a phase. For a single qutrit, the Pauli group $\mathcal{P}_{3}$ is generated by X and Z gates, while the minimal generating set for the Clifford group $\mathcal{C}_{3}$ is given by H and S gates \cite{morvan2021qutrit,ringbauer2022universal}, in which the Z gate and the S gate  are respectively defined as
\begin{equation}
\mathrm{H} = \frac{1}{\sqrt{3}}\begin{pmatrix}
 1 & 1 & 1\\
 1 & \omega  &\omega^{2} \\
 1 & \omega^{2} & \omega 
\end{pmatrix}, \quad
\mathrm{X} = \begin{pmatrix}
 0 & 0 & 1\\
 1 & 0  &0 \\
 0 & 1 & 0 
\end{pmatrix}, \quad
\mathrm{Z} = \begin{pmatrix}
 1 & 0 & 0\\
 0 & \omega  &0 \\
 0 & 0 & \omega^{2} 
\end{pmatrix}, \quad
\mathrm{S} = \begin{pmatrix}
 1 & 0 & 0\\
 0 & \omega  &0 \\
 0 & 0 & 1
\end{pmatrix}.
\end{equation}
Although the minimal set $\{\mathrm{H}, \mathrm{S}\}$ is mathematically elegant and sufficient to generate all single-qutrit Clifford operations, it is not enough efficient in experimental settings. For instance, the qutrit X gate, which is commonly used in quantum algorithms as a Pauli gate, is itself a Clifford gate. However, constructing it from the minimal set requires a long gate sequence, viz.,
\begin{equation}
 \mathrm{X}  = \mathrm{H} \cdot \mathrm{S}^2 \cdot \mathrm{H}^{2} \cdot \mathrm{S} \cdot \mathrm{H} ,
\end{equation}
which results in significant cumulative errors due to increased circuit depth. From an experimental perspective, reducing circuit depth is crucial. In general, generating a Clifford gate requires, on average, 
4.06 H gates and 4.89 S gates
with the minimal generating set. In this work, we leverage the redundant generating set $\{\mathrm{H}, \mathrm{S} ,\mathrm{X}, \mathrm{Z} \}$ to implement single-qutrit Clifford operations \cite{liu2023performing}.
We provide an example pseudocode as shown in Alg. \ref{alg:su3_clifford}.
This yields an average gate count of
1.542 H gates, 1.625 S gates, 0.125 X gates, and 1.648 Z gates
per Clifford operation, offering a substantially more efficient and practical alternative.

Since both S and Z gates can be realized virtually as phase gates, combining them with H and X gates enables the efficient implementation of the single-qutrit Clifford group $\mathcal{C}_{3}$ . By further introducing a non-Clifford T gate, defined as the qutrit analogue of the qubit $\pi/8$ gate \cite{howard2012qudit}, denoted as
\begin{equation}
\mathrm{T} = \begin{pmatrix}
 1 & 0 & 0\\
 0 & e^{\frac{2\pi i}{9} }  &0 \\
 0 & 0 & e^{\frac{-2\pi i}{9} } 
\end{pmatrix} ,
\end{equation}
which can be also realized virtually, we obtain a universal gate set for single-qutrit gate operations. As a result, any qutrit unitary gate can be approximated to arbitrary precision by a finite sequence of these gates, as guaranteed by the Solovay–Kitaev theorem extended to qutrits \cite{ringbauer2022universal,prakash2018normal}.

\subsection{Comparison with the SU(2) Decomposition}

In this section, we provide a detailed comparison between our direct SU(3)-based control framework and the conventional SU(2) decomposition schemes from three complementary perspectives: (i) the compilation of arbitrary SU(3) operations, (ii) the synthesis of the qutrit Clifford group, and (iii) the implementation of the qutrit H gate.

\textbf{(i) Decomposition of arbitrary SU(3) operations.} 
Conventional SU(2)-based schemes employ the Givens decomposition, which generalizes the Euler decomposition of SU(2) to higher dimensions \cite{lindon2023complete}. In its standard form, an arbitrary $U \in \mathrm{SU}(3)$ can be expressed as a product of three SU(2) rotations acting on different subspaces of the three-level Hilbert space. 
\begin{equation}
U = U_{12}(\theta_3, \phi_3, \lambda_3)\,
    U_{01}(\theta_2, \phi_2, \lambda_2)\,
    U_{12}(\theta_1, \phi_1, \lambda_1),
\end{equation}
where $U_{ij}(\theta_k, \phi_k, \lambda_k)$ denotes a unitary operation acting only on the two-dimensional subspace 
$\{\ket{i}, \ket{j}\}$, while leaving the other level unchanged.
Within this subspace, the general form of $U_{ij}(\theta_k, \phi_k, \lambda_k)$ can be expressed as
\begin{equation}
U_{ij}(\theta_k, \phi_k, \lambda_k) =
\begin{pmatrix}
\cos \frac{\theta_k}{2} & -i e^{i\lambda_k} \sin \frac{\theta_k}{2} \\
-i e^{i\lambda_k} \sin \frac{\theta_k}{2} & e^{i(\phi_k+\lambda_k)} \cos \frac{\theta_k}{2}
\end{pmatrix}_{(ij)}.
\end{equation}
For each of these two-levels subspace, we use the so-called U3 decomposition to implement arbitrary rotation in these subspace, that is, $U_{ij}(\theta_k, \phi_k, \lambda_k) = \text{Z}^{(ij)}_{\phi_k-\pi} \text{H}^{(ij)} \text{Z}^{(ij)}_{- \theta_k} \text{H}^{(ij)} \text{Z}^{(ij)}_{\lambda_k - \pi}$,
where the suberscript $(ij)$ indicates the corresponding subspace \cite{morvan2021qutrit}. As a result, each rotation requires 2 native $\pi/2$ ($\text{H}^{(01)}$ or $\text{H}^{(12)}$) pulses. Thus, at most 6 native pulses ( excluding software-defined phase gates) are required to realize a general SU(3) operation. 

However, we are not aware of any existing studies that explicitly construct arbitrary SU(3) unitaries exclusively from native qutrit gates. Here, following SU(2) decomposition-based scheme, we propose a similar sequence using alternating phase and H gates: 
\begin{equation}\label{eq:SU3_decomposition}
U = \text{Z}_{\phi_1} \text{H} \text{Z}_{\phi_2} \text{H} \text{Z}_{\phi_3} \text{H} \text{Z}_{\phi_4} \text{H} \text{Z}_{\phi_5},
\end{equation}
where the generalized phase gate is defined as $\text{Z}{\phi_j} = \mathrm{diag} (1, e^{i\phi^{(1)}_j}, e^{i(\phi^{(1)}_j + \phi^{(2)}_j) })$.
Intuitively, the sequence contains ten independent parameters, which provide sufficient degrees of freedom to explore the eight-dimensional manifold of $\mathrm{SU}(3)$. 
Although a rigorous mathematical proof of universality is nontrivial and beyond the scope of this work, the completeness of the proposed sequence can be supported by both Lie algebra viewpoint and numerical evidence. 
From the Lie algebra  perspective, $\mathfrak{su}(3)$ consists of eight generators, i.e. two diagonal and six off-diagonal elements. 
The diagonal phase gates $\text{Z}_{\phi_i}$ generate the two diagonal elements, while the non-diagonal H gate mixes the basis states, and through successive commutators with $\text{Z}_{\phi_i}$, can generate all off-diagonal elements.
Together, these operators span the full $\mathfrak{su}(3)$ algebra and thus the entire $\mathrm{SU}(3)$ manifold.
Meanwhile, we provide a pseudocode for numerical calculation, as shown in Alg. \ref{alg:su3_decomposition}, to illustrate the flexibility of the proposed decomposition Eq. \eqref{eq:SU3_decomposition}.

Wihtout loss of generality, we numerically tested a randomly generated SU(3) unitary defined as
\begin{equation}
U_{\mathrm{rand}} =
\begin{pmatrix}
-0.20279349-0.85979105i & 0.05722587+0.08779234i & -0.43138164-0.15020568i \\
-0.19776429+0.03209117i &  -0.26085299-0.82004731i & -0.00944768-0.46823959i \\
0.05121576+0.4205598i & 0.49080444+0.0872172i & -0.63803545-0.40610506i
\end{pmatrix}.    
\end{equation}
Through numerical optimization, we find that the parameter set 
$\phi_1 = (3.833850, 6.093237)$, $\phi_2 = (4.564811, 3.185648)$, $\phi_3 = (1.452682, 4.539998)$, $\phi_4 = (3.182307, 4.601076)$ and $\phi_5 = (3.871586, 0.018866)$
can accurately reproduce the target unitary with an error below $10^{-9}$
under a global phase $\phi_{0} = 2.7556666$ defined as $e^{-i \phi_0}$.
It is note that, here, the gate error is routinely defined as $\varepsilon(U_{\mathrm{rand}},U) = 1 - \big|\mathrm{Tr}(U_{\mathrm{rand}}^{\dagger}U)/d\big|^{2}$ with the Hilbert space dimension $d=3$, such that the global phase is irrelevant.
This example provides numerical evidence that the proposed sequence is feasible to realize arbitrary SU(3) operations.

As a result, our scheme reduces the number of native pulses by approximately $1/3$ compared with the conventional
SU(2) decomposition. Moreover, by including X or $\text{H}^{-1}$ gates, as demonstrated in the main text, which are readily implemented using our scheme, the circuit depth may be further reduced.

\textbf{(ii) Decomposition of the qutrit Clifford group.}
Note that although, from an experimental perspective, the universality of the Clifford+T gate set is only meaningful in the fault-tolerant regime, where both Clifford and non-Clifford T gates can be implemented at the logical level with quantum error correction. The conceptual implications of the Solovay–Kitaev theorem extend far beyond that context, which highlights that high-fidelity physical Clifford gates serve as the essential building blocks of circuit-based quantum computation, delineating a practical pathway to fault-tolerant quantum computation. 
In fault-tolerant architectures such as the surface code, logical Clifford operations are directly constructed from accurate physical Clifford gates.  
Meanwhile, the non-Clifford T gate, which is not naturally compatible with stabilizer codes, is realized through magic-state distillation, a process that itself relies on Clifford operations and post-selection to purify noisy resource states into the high-fidelity magic states required for T gate execution.  
Therefore, achieving high-fidelity Clifford gates on physical qudits, as demonstrated in our work, represents a crucial step toward fault-tolerant qudit-based quantum computation.

In the following, we compare the average number of native pulses required for a single qutrit Clifford operation across different schemes. 
For the SU(2) decomposition using only $\pi/2$ pulses ($\text{H}^{(01)}$ and $\text{H}^{(12)}$ gates), the average Clifford gate requires 5.25 native pulses \cite{liu2023performing}. 
When both $\pi/2$ pulses and $\pi$ pulses ($\text{X}^{(01)}$, $\text{X}^{(12)}$ gates) are allowed, the number reduces to 3.325 per Clifford \cite{morvan2021qutrit}. 
In contrast, our direct SU(3)-based scheme requires only 1.667 native gates on average, and may be further reduced when including arbitrary phase gates. 
Our scheme therefore achieves a substantial reduction in Clifford compilation depth, a 68.2\% improvement over the basic $\pi/2$-pulse SU(2) decomposition and a 49.9\% improvement over an optimized approach that utilizes both $\pi/2$ and $\pi$ pulses.

\textbf{(iii) Comparison of Hadamard gate implementations.}
The H gate is of paramount importance in quantum computing due to its dual role as a Clifford group generator and a key building block for major quantum algorithms, most notably the qutrit-based Deutsch-Jozsa, Grover, and Bernstein-Vazirani algorithms \cite{wang2020qudits}. This has motivated extensive experimental efforts to realize it with high fidelity.
Table~\ref{tab:compare} summarizes the performance of our single-step H-gate realization in comparison with recent SU(2)-based implementations.
For a fair comparison, all fidelities are converted to process fidelities \cite{morvan2021qutrit}.

\begin{table}[h]
\centering
\caption{Comparison of qutrit H gate implementations.}
\label{tab:compare}
\setlength{\tabcolsep}{10pt}
\renewcommand{\arraystretch}{1.2}
\begin{tabular}{lcccc}
\hline\hline
Reference & Method & Duration (ns) & Fidelity (\%) & $T_1^{01/12}$, $T_2^{01/12}$ ($\mu$s) \\
\hline
Morvan \textit{et al.} (2021) \cite{morvan2021qutrit}& SU(2) & 124 & 99.4 & 60/36, 60/27 \\
Roy \textit{et al.} (2023) \cite{roy2023two} & SU(2) & 142 & 97.5 & 48/22, 4.5/2.0 \\
Liu \textit{et al.} (2023) \cite{liu2023performing} & SU(2) & 190 & 98.4 & 180/101, 76/37 \\
\textbf{Our work (2025)} & \textbf{SU(3)} & \textbf{35} & \textbf{99.4} & 61/28, 4.6/4.4 \\
\hline\hline
\end{tabular}
\end{table}
As discussed in Sec.~\ref{sec:s3_cali}, with careful calibration, the dominant errors in our SU(3)-based single-step protocol stem from decoherence processes, particularly pure dephasing. Despite this limitation, our experimentally demonstrated H gate achieves a fidelity comparable to or even higher than SU(2)-based decomposition schemes while simultaneously reducing the gate duration by a factor of 3-5, as indicated in Table~\ref{tab:compare}.
This result clearly highlights the advantage of our scheme in gate duration, which is a direct consequence of reducing circuit depth.
As coherence times continue to improve, our approach has the potential to achieve single-qutrit gate fidelities comparable to those of state-of-the-art single-qubit gates, demonstrating its promise for future high-performance qutrit processors.

\section{Quantum circuits}
In the main text, we have demonstrated two quantum circuits implemented with a single qutrit. In fact, both circuits can be extended to the qudit case with $d > 3$. In the following, we generalize these two quantum circuits to the higher-dimensional qudit and provide a comprehensive discussion of their potential applications. Note that for a qudit, the $\mathrm{H}_{d}$ gate is actually a form of the discrete Fourier transform matrix \cite{ringbauer2022universal,prakash2018normal}, that is,
\begin{equation}
\begin{aligned}
\mathrm{H}_{d} &= \frac{1}{\sqrt{d}} \sum_{j,k=0}^{d-1} e^{2\pi i jk / d} \, |j\rangle\langle k|       
\\
&= \frac{1}{\sqrt{d}} \begin{pmatrix}
1 & 1 & 1 & \cdots & 1 \\
1 & \omega & \omega^2 & \cdots & \omega^{d-1} \\
1 & \omega^2 & \omega^4 & \cdots & \omega^{2(d-1)} \\
\vdots & \vdots & \vdots & \ddots & \vdots \\
1 & \omega^{d-1} & \omega^{2(d-1)} & \cdots & \omega^{(d-1)^2}
\end{pmatrix}
\end{aligned}
\end{equation}
where $\omega = e^{2\pi i / d}$ is the primitive $d$-th root of unity. 

\subsection{Ramsey interferometer}
We consider a qudit initialized in the ground state $\ket{0}$, , which is subjected to a Ramsey interferometric sequence consisting of a $\mathrm{H}_{d}$ gate, followed by a phase gate $\mathrm{Z}_{d,\phi}$, and concluded with the inverse gate $\mathrm{H}_{d}^{-1}$. The qudit phase gate is defined as $\mathrm{Z}_{d,\phi} = \mathrm{diag}(1,e^{i\phi},e^{2i\phi},...,e^{i(d-1)\phi})$, representing free evolution characterized by a phase parameter $\phi$. The final state of the system is given by
\begin{equation} \label{eqs:31}
\begin{aligned}
\ket{\psi (\phi)} &= \mathrm{H}_{d}^{-1} \mathrm{Z} _{d,\phi} \mathrm{H}_{d} \ket{0} \\
&= \frac{1}{d}\sum_{j=0}^{d-1} \frac{1-e^{id(\phi-2\pi j/d)} }{1-e^{i(\phi-2\pi j/d)}} \ket{j} .
\end{aligned}  
\end{equation}
Consequently, the population $P_{k}(\phi) $ in state $\ket{k}$, ($k = 0,1,...,d-1$) as a function of $\phi$ can be expressed in a form analogous to that of grating interference,
\begin{equation}
P_{k}(\phi) = \frac{1}{d^2} \frac{\sin^2 \left( \frac{d}{2} (\phi - \frac{2\pi k}{d}) \right)}{\sin^2 \left( \frac{1}{2} (\phi - \frac{2\pi k}{d}) \right)},
\end{equation}
This expression corresponds to a standard Dirichlet kernel, exhibiting periodic interference fringes akin to a linear phase comb. We numerically simulated the interference fringes for $d = 2, 3, 4, 5$, with the results presented in Fig.~\ref{fig:s10}.

\renewcommand{\thefigure}{S\arabic{figure}} 
\begin{figure}[ht]
    \centering 
    \includegraphics[width = 0.80\textwidth, height = 0.25\textheight]{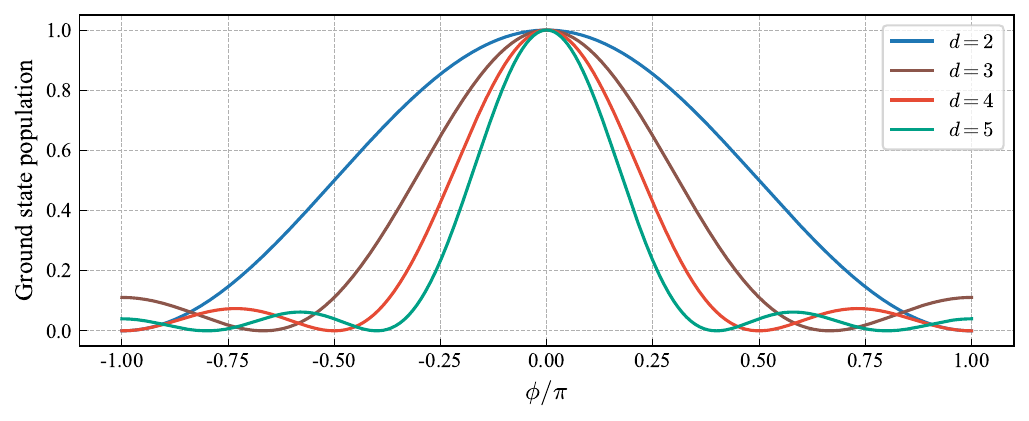}
    \caption{\justifying Qudit ($d$ = 2, 3, 4, 5) Ramsey interference patterns showing the ground state population as a function of the delay phase $\phi$, with the qudit initially prepared in state $\ket{0}$. The population of state $\ket{j}$ follows a similar grating interference pattern, shifted by a phase of $2\pi j/d$ relative to state $\ket{0}$.}
    \label{fig:s10}
\end{figure}

The Ramsey interferometer is a well-established experimental setting for phase estimation. Just as a diffraction grating offers enhanced phase sensitivity compared to a double-slit setup, the qudit-based Ramsey interferometer is expected to outperform its qubit counterpart high-precision phase metrology. The phase  measurement precision can be quantified by \cite{yu2022quantum}
\begin{equation}
\delta \phi =\frac{\Delta P}{\partial P/\partial \phi} , 
\end{equation}
where we estimate the phase by the projective measurement $P = \ket{0} \bra{0}$ in the ground state, without loss of generality. Here $\Delta P$ denotes the standard derivative associated with the measurement result, which decreases as $1/\sqrt{N}$ with the number of individual measurements $N$ due to statistical averaging. As shown in Fig.~\ref{fig:s11}(a), the phase estimation precision for a single measurement exhibits a strong dependence on the phase $\phi$ for various qudit dimensions $d = 2, 3, 4, 5$, with the optimal precision consistently attained at $\phi = 0$. Note that the optimal precision attained at $\phi = 2\pi j/d$ if the projective measurement is performed on state $\ket{j}$.

In addition, the single-measurement precision of any unbiased estimator is fundamentally bounded by the inverse of the quantum Fisher information (QFI) $F_{\phi}$, as dictated by the quantum Cram\'{e}r–Rao bound \cite{Toth2014,yu2022quantum}
\begin{equation}
\delta \phi \ge 1/\sqrt{F_{\phi}} ,
\end{equation}
where the QFI is defined as $F_{\phi} = 4[\bra{\partial_{\phi} \psi (\phi)} \partial_{\phi} \psi (\phi) \rangle - | \bra{\psi (\phi)} \partial_{\phi} \psi (\phi) \rangle|^{2} ]$, quantified the sensitivity of the quantum state to changes in the parameter $\phi$. For the qudit Ramsey interferometer under consideration, the QFI can be calculated explicitly as $F_{\phi} = (d^{2}-1)/3$ in terms of Eq.~(\ref{eqs:31}), consistent with the quantum Cram\'{e}r–Rao bound, as shown in Fig.~\ref{fig:s11}(b). 

\renewcommand{\thefigure}{S\arabic{figure}} 
\begin{figure}[ht]
    \centering 
    \includegraphics[width = 0.90\textwidth, height = 0.26\textheight]{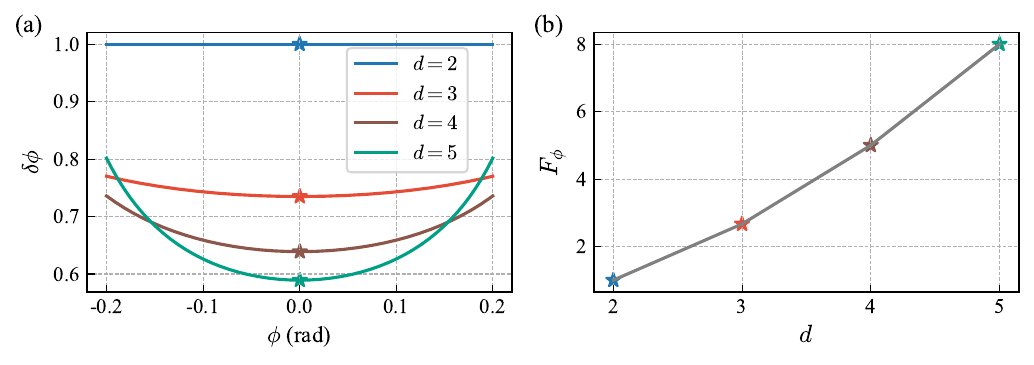}
    \caption{\justifying (a) Phase measurement precision $\delta\phi$ as a function of phase shift $\phi$ for qudit dimensions $d = 2, 3, 4, 5$. The minimum of each curve is marked with a star, corresponding to the optimal working point. (b) Quantum Fisher information $F_{\phi}$ as a function of qudit dimension $d$. Stars denote the values for $d = 2, 3, 4, 5$, matching the optimal sensitivities indicated in (a), where the Cram\'{e}r–Rao bound is saturated. The numerical results indicate the metrological advantage of higher-dimensional qudits for phase estimation, since increased dimension yields higher measurement precision and greater Fisher information.}
    \label{fig:s11}
\end{figure}

While QFI merely provides a theoretical upper bound on estimation precision, the Kitaev algorithm furnishes a concrete phase estimation protocol, which sequentially determines the binary digits encoding the target phase through iterated Ramsey interference with varied phase delays \cite{shlyakhov2018quantum,danilin2018quantum}. Here, we develop a qudit generalization for the Kitaev algorithm by extending the conventional binary processing to a base-$d$ representation, significantly enhancing phase estimation performance. Specifically, without loss of generality, we parameterize the target phase as $\phi = 2\pi \theta$, where the normalized phase $ \theta \in [0, 1)$ admits a base-$d$ expansion that provides $N$-digit precision written as
\begin{align}
\phi_{d,N}/2\pi  &= \frac{\theta_{0} }{d} + \frac{\theta_{1} }{d^2}+...+\frac{\theta_{N-1} }{d^N}  \\
&= 0.\theta_{0} \theta_{1}... \theta_{N-1}, \quad (\mathrm{base}-d)
\end{align}
where $\theta_{k} \in \{0,1,...,d-1 \}$ are the $d$-ary digits. As we shall demonstrate, all $N$ $d$-ary digits can be sequentially determined through $N$ iterative Ramsey interference cycles, proceeding from the least significant digit $\theta_{N-1}$ to the most significant digit $\theta_{0}$. Each cycle employs distinct phase delays to extract the corresponding digit.
Notably, the phase estimation algorithm can be employed in flux sensors or magnetometers based on the transmon platform, owing to the flux-dependent transition frequencies of the transmon qudit, given by $\omega_{n} = n\omega_{1}(\Phi) +  \frac{n(n-1)}{2} \alpha$, where $\alpha$ denotes the fixed anharmonicity. In particular, the linear energy levels associated with the flux-dependent component facilitate the straightforward implementation of $Z_{d,\phi}$ gates, with phase delays precisely controlled via the duration of free evolution.

In the Kitaev algorithm, the phase delay of the intermediate phase gate in the Ramsey interferometer is first set to $d^{N-1} \phi$, where $d^{N-1} \phi = (2\pi /d) \theta_{N-1}$ mod $2\pi$. Then it can be verified that the final state coincides with one of the computational basis states, depending on the value of $\theta_{N-1}$. In particular, final state $\ket{m}$ corresponds to $\theta_{N-1} = m$. Therefore, the measurement outcome unambiguously determines the least significant base-$d$ digit $\theta_{N-1}$. Next, the phase delay in the Ramsey interferometer is reduced to $1/d$, yielding $d^{N-2} \phi =  (2\pi /d) \theta_{N-2} + (2\pi /d^{2}) \theta_{N-2} $ mod $2\pi$.
Note that the latter can be eliminated using the value of $\theta_{N-1}$ from the previous measurement. The subsequent projective measurement thus provides the next base-$d$ digit, $\theta_{N-2}$, in a deterministic way. Proceeding analogously with gradually decreased phase delays $d^{N-3} \phi, d^{N-4} \phi,..., \phi$ allows for the unambiguous determination of all base-$d$ digit.
Here, the values $(2\pi /d) \theta_{j}$ , where $\theta_{j} \in \{0, 1,..., d-1 \} $, correspond precisely to the optimal measurement points that saturate the quantum Cram\'{e}r–Rao bound, hence allowing a single measurement more precisely.

For a target phase $\phi$, the normalized probability density of obtaining the measurement outcome $\tilde{\phi}$ after $N$ iterations of Ramsey interference is given by
\begin{align}
p(\tilde{\phi}|\phi ) &\propto  \prod_{k=0}^{N-1} \frac{\sin^{2}(d^{k}  \delta \phi /2)}{d^{2}\sin^{2}( d^{k}\delta \phi /2) } \\
&= \frac{1}{2\pi d^{N}}  \frac{\sin^{2}(d^{N}  \delta \phi /2)}{\sin^{2}( \delta \phi /2)}.
\end{align}
As shown in Fig.~\ref{fig:s12}, we present the probability density distributions as functions of the iteration number and the qudit dimension. As the iterations $N$ or the dimension $d$ increases, the distribution becomes progressively sharper, with the width of the main peak scaling as $1/d^{N}$. As a result, the qudit Ramsey interferometer allows to extract more phase information per iteration, reducing the number of individual measurements required by a factor of $O(\log{d})$ compared to the qubit case to achieve a given degree of phase precision.

\renewcommand{\thefigure}{S\arabic{figure}} 
\begin{figure}[ht]
    \centering 
    \includegraphics[width = 0.80\textwidth, height = 0.24\textheight]{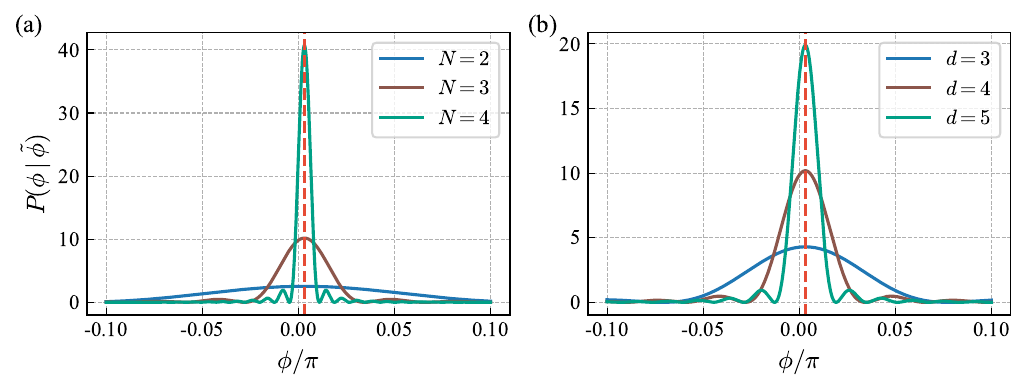}
    \caption{\justifying Probability density distribution in the Kitaev algorithm: (a) dependence on the iteration number for $d=4$ and (b) dependence on the qudit dimension for $N=3$. In both panels, the target phase is fixed at $\tilde{\phi}=0.01\pi$, as indicated by the vertical dashed line.}
    \label{fig:s12}
\end{figure}

\subsection{Parity-check algorithm}
We have experimentally demonstrated a parity-check algorithm for the three-element symmetric group $S_{3}$ using a transmon qutrit, based on our proposed gate-design protocol. As discussed in the main text, this algorithm can be extended to general qudit ($d \ge 3$) case for determining the parity of all $2d$ elements in the dihedral group $D_{d}$, which corresponds to the symmetry group of a regular $d$-gon and constitutes a subgroup of the $2d$-element symmetric group $S_{d}$ \cite{gedik2015computational}. Noted that the isomorphism $S_{3}\simeq D_{3}$ is a special case. Although this algorithm has been addressed in the previous work, we present here a more rigorous and formalized treatment based on group theory.

In the parity-check algorithm, a qudit initialized in the state $\ket{m}$ is transformed into the Fourier basis $\ket{f_{m}} = \frac{1}{\sqrt{d}}{\textstyle \sum_{k=0}^{d-1}}  \chi_{m}(k) \ket{k}$ via the $H_{d}$ gate, where $\chi_{m}(k) = \omega^{mk}$ denotes an irreducible character of the abelian group $\mathbb{Z}_{d}$ when $\gcd(m,d)=1$. A unitary operator $U_{\pi}$, representing the action of a permutation $\pi \in D_{d}$, is then applied. This operator maps permutations in the computational basis to corresponding transformations in the Fourier basis. Finally, the inverse gate $H_{d}^{-1}$ is applied to return to the computational basis prior to measurement. The resulting final state $\ket{\psi_{f}}$ can then be expressed as
\begin{align}
\ket{\psi_{f}} &= H_{d}^{\dagger}U_{\pi}H_{d}\ket{m} \\
&=\frac{1}{d} \sum_{n=0}^{d-1} \bra{f_{n} } U_{\pi } \ket{f_{m} } \ket{n} \\
&= \frac{1}{d}\sum_{n=0}^{d-1} \sum_{k=0}^{d-1} \chi_{m}(k)  \chi_{n}^{\ast } (\pi(k))  \ket{n} .
\end{align}

When $\pi$ is an even permutation, it can be represented as a cyclic shift by $r$, i.e., $\pi(k) = k + r \mod d$. The character then transforms as $\chi_n(\pi(k)) = \omega^{n(k + r)} = \omega^{nr} \chi_n(k)$. For an odd permutation of the form $\pi(k) = -k + r \mod d$, the character becomes $\chi_n(\pi(k)) = \omega^{n(-k + r)} = \omega^{nr}\chi_{-n}(k)$. Utilizing the orthogonality of the irreducible characters of $\mathbb{Z}_{d}$, the final state satisfies
\begin{equation}
\ket{\psi_{f}} = 
\begin{cases} 
\ket{m} & \text{for even permutation,}  \\ 
\ket{d-m} & \text{for odd permutation.} 
\end{cases}
\end{equation}
As a result, for suitable choices of the initial state $\ket{m}$, with $\gcd(m,d)=1$, even and odd permutations yield orthogonal final states, enabling deterministic parity discrimination \cite{liu2023performing}. In particular, the state $\ket{m}$ remains unchanged under an even permutation, while it transforms into $\ket{d-m}$ under an odd permutation. As illustrated in Fig.~\ref{fig:s13}, we simulate several representative permutations from $D_{5}$ with initial state set as $\ket{2}$. Since a classical algorithm requires querying at least twice to determine the parity of a permutation in $D_{d}$, one to identify the cyclic shift and another for its direction, the quantum algorithm  allows for parity determination with a single query, achieving a two-to-one speedup without entanglement.

\renewcommand{\thefigure}{S\arabic{figure}} 
\begin{figure}[ht]
    \centering 
    \includegraphics[width = 0.70\textwidth, height = 0.35\textheight]{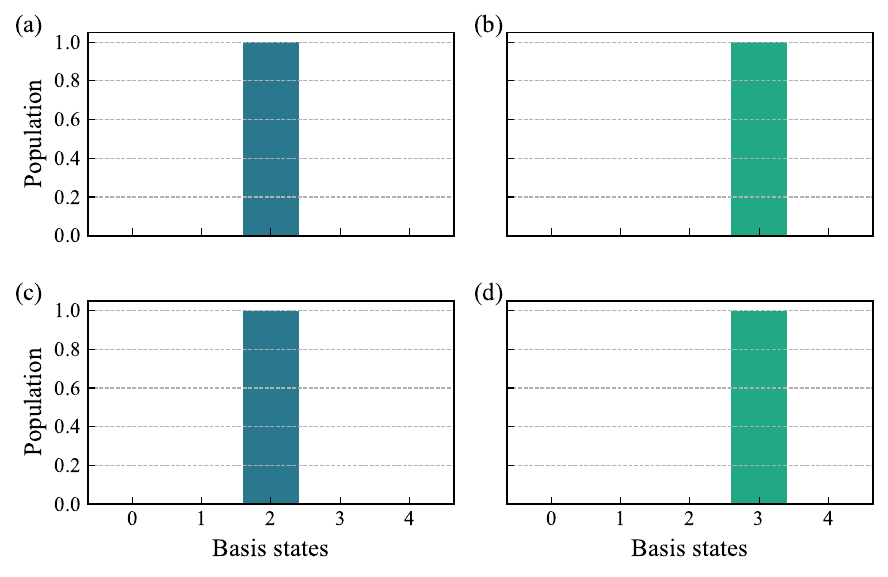}
    \caption{\justifying Population distribution in the computational basis states after performing the parity-check algorithm for the permutations (a) (12340), (b) (43210), (c) (23401), and (d) (32104), with the qudit initialized in state $\ket{2}$. For the even permutations (a) and (c), the final state remains $\ket{2}$, while for the odd permutation (b) and (d), the final state transitions to $\ket{3}$.}
    \label{fig:s13}
\end{figure}

\section{Sample information}
\subsection{Coherence characterization}
Our qutrit platform is realized on a superconducting transmon featuring a multi-level structure, with the lowest three energy levels $ \{ \ket{0}, \ket{1}, \ket{2} \}$ as the computational subspace. The transmon consists of a cross-shaped grounded capacitance and a superconducting quantum interference device (SQUID) loop, which incorporates two symmetric Josephson junctions. It is inductively coupled to a control line to realize Z manipulation for frequency biasing and an XY manipulation for introducing microwave drives. Meanwhile, the transmon is also capacitively coupled to a quarter-wavelength resonator, which is further coupled to a transmission line, forming a circuit QED architecture that enables dispersive readout of the qubit state.

\renewcommand{\arraystretch}{1.2} 
\begin{table}[htbp]
    \centering
    \caption{Device characterization}
    \label{Table_1}
    \setlength{\tabcolsep}{5mm}{
    \begin{tabular}{cccc}
       \toprule
       {Transitions} & {$ \ket{0} \leftrightarrow \ket{1}  $} & {$ \ket{1} \leftrightarrow \ket{2} $} &{$ \ket{0} \leftrightarrow \ket{2} $} \\
        \midrule
        $\omega /2\pi$ (GHz) & 4.993 & 4.800 & 4.896 \\ 
        $T_{1}(\mu s)$ & 60.7 & 28.4 & 523.1 \\
        $T_{2,R}(\mu s)$ & 4.6 & 4.4 & 4.2 \\
        \bottomrule
    \end{tabular}}
\end{table}

The experiment is performed with the transmon qutrit biased at its flux-insensitive sweet spot, minimizing susceptibility to flux noise. To fully characterize the qutrit's coherence properties, we conducted time-domain measurements on all three relevant transitions: $\ket{0} \leftrightarrow \ket{1}$, $\ket{1} \leftrightarrow \ket{2}$, and the two-photon transition $\ket{0} \leftrightarrow \ket{2}$. The corresponding transition frequencies and coherence times are summarized in Table~\ref{Table_1}. Note that the transmon qutrit exhibits a relatively weak anharmonicity of approximately 193 MHz, which leads to substantial cross-coupling effects that manifest as dominant coherent errors during gate operations. Therefore, a series of calibration procedures is required to mitigate such effects. Details of the calibration protocol are described in Sec.~\ref{sec:s3_cali}.

Energy relaxation times $T_{1}$ are extracted by initializing the system in either $\ket{1}$ or $\ket{2}$ and monitoring the population decay across all three levels. A global fit is applied using a set of coupled rate equations, viz.,
\begin{equation} \label{eqs:28}
\begin{aligned}
\frac{dP_{2}}{dt} &= - \frac{1}{T_{1}^{(12)}} P_{2}(t) + \frac{1}{T_{1}^{(02)}}  P_{2}(t) ,  \\
\frac{dP_{1}}{dt} &= + \frac{1}{T_{1}^{(12)}} P_{2}(t) - \frac{1}{T_{1}^{(01)}} P_{1}(t),  \\
\frac{dP_{0}}{dt} &= + \frac{1}{T_{1}^{(01)}} P_{1}(t) + \frac{1}{T_{1}^{(02)}} P_{2}(t),
\end{aligned}
\end{equation}
which accounts for both the cascade decay $\ket{2} \to \ket{1} \to \ket{0}$ and direct decay $\ket{2} \to \ket{0}$ \cite{peterer2015coherence}. This method enables the accurate determination of the relaxation times $T_{1}^{(01)}$, $T_{1}^{(12)}$, and $T_{1}^{(02)}$, as shown in Fig.~\ref{fig:s5}. To probe dephasing, we apply Ramsey sequences that prepare superposition states $(\ket{0}+\ket{1})/\sqrt{2}$, $(\ket{1}+\ket{2})/\sqrt{2}$, and $(\ket{0}+\ket{2})/\sqrt{2}$ for the respective transitions. Ramsey dephasing times $T^{(01)}_{2,R}$, $T^{(12)}_{2,R}$ and $T^{(02)}_{2,R}$ are extracted by fitting the resulting dynamics to a model incorporating the known $T_{1}$ contributions from Eqs.~(\ref{eqs:28}). The coherence decay is modeled using a stretched exponential, $f(t) = \exp[-(t/T_2)^n]$, to capture the noise spectral characteristics, as shown in Fig.~\ref{fig:s6}. Across all transitions, the measured $T_{2,R}$ values are considerably shorter than the corresponding $T_{1}$ limits, with stretching exponents $n \approx 1$, indicating that coherence is predominantly limited by pure dephasing due to Lorentzian-like noise.

\renewcommand{\thefigure}{S\arabic{figure}} 
\begin{figure}[ht]
    \centering
    \includegraphics[width = 0.80\textwidth, height = 0.22\textheight]{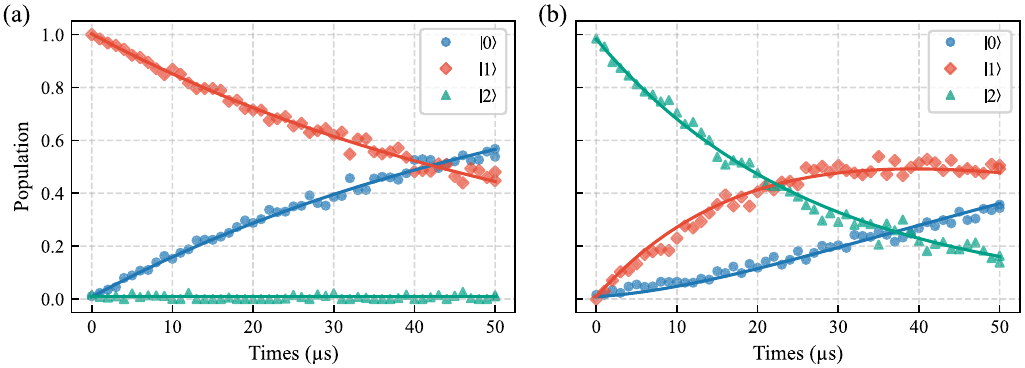}
    \caption{\justifying Population decay dynamics with the qutrit initialized in (a) $\ket{1}$ and (b) $\ket{2}$, respectively. The solid lines represent fits to a rate-equation model that accounts for all decay channels, enabling the extraction of energy relaxation times $T^{01}_{1}$, $T^{12}_{1}$, and $T^{02}_{1}$.}
    \label{fig:s5}
\end{figure}

\renewcommand{\thefigure}{S\arabic{figure}} 
\begin{figure}[ht]
    \centering
    \includegraphics[width = 0.92\textwidth, height = 0.20\textheight]{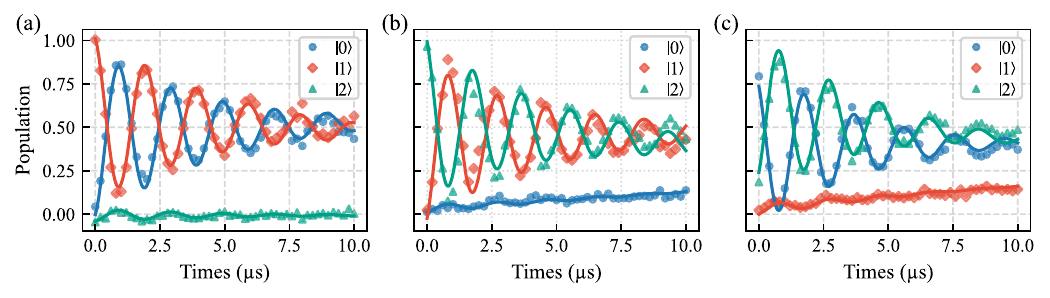}
    \caption{\justifying Ramsey oscillations for the transitions (a) $\ket{0} \leftrightarrow \ket{1}$, (b) $\ket{1} \leftrightarrow \ket{2}$, and (c) $\ket{0} \leftrightarrow \ket{2}$, respectively. The solid lines are fits to a Ramsey interference model incorporating stretched-exponential coherence decay and energy relaxation described by a rate-equation approach, from which the Ramsey dephasing times $T^{01}_{2,R}$, $T^{12}_{2,R}$, and $T^{02}_{2,R}$ are extracted.}
    \label{fig:s6}
\end{figure}

\subsection{Qutrit state readout}
The state of the transmon can be inferred from the state-dependent dispersive shift of the readout resonator frequency. In particular, this is achieved by driving the resonator with a readout pulse of approximately 1 $\mu$s duration and subsequently measuring the transmission coefficient $S_{21}$, which can be represented as a point within the $I-Q$ space. In our experimental setup, the single-shot readout is limited, particularly for higher energy levels as a result of their shorter coherence times. Here, we have developed an improved method for accurately determining state populations, which extends the conventional voltage-to-population conversion techniques used in qubit systems by leveraging the temporal stability of measured response voltages. Instead of relying solely on a single readout frequency, our method employs multiple optimized readout frequencies to enhance the accuracy of population estimations. Specifically, as illustrated in Fig.~\ref{fig:s7}, for the case of a qutrit, we fix the readout frequency at three resonance points, each corresponding to one of the computational basis states. At any given time, we repeatedly measure the response voltages $V_{j}$ $(j=1,2,3)$ at each frequency point $j$, which are related to the state populations through the linear equation
\begin{equation}
V_{j}= P_{0} V_{0,j} + P_{1} V_{1,j} +P_{2} V_{2,j},
\end{equation}
where $V_{n,j}$ $(n=0,1,2)$ denotes the characteristic response voltage obtained when the system is prepared on the state $\ket{n}$ at frequency point $j$. Thus, by solving this set of linear equations, the population $P_{n}$ associated with each basis state $\ket{n}$ can be accurately reconstructed. It is worth noting that through carefully selecting the optimal readout frequency, or incorporating additional measurement points with solving the resulting over-determined results with optimization algorithms, the standard deviation of the population estimations can potentially be further reduced.

\renewcommand{\thefigure}{S\arabic{figure}} 
\begin{figure}[ht]
    \centering
    \includegraphics[width = 0.70\textwidth, height = 0.26\textheight]{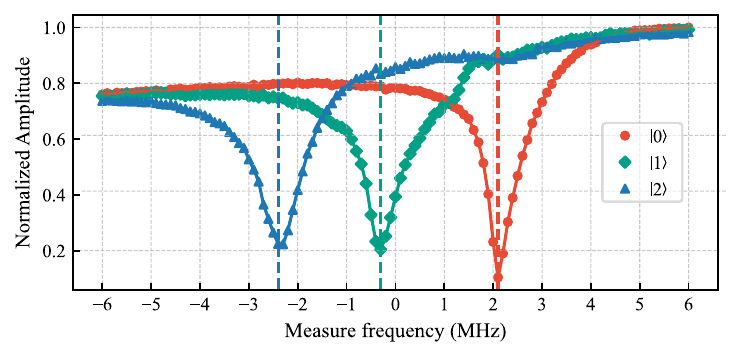}
    \caption{\justifying Voltage response as a function of measurement frequency. The resonator exhibits distinct responses depending on whether the qubit is prepared in state $\ket{0}$, $\ket{1}$ or $\ket{2}$. The vertical dashed line indicates the frequency chosen for state discrimination.}
    \label{fig:s7}
\end{figure}

\section{Calibration procedures} \label{sec:s3_cali}

\subsection{Error analysis}

Quantum gates are subject to both incoherent and coherent errors. The incoherent error for a single-qutrit gate implemented by a short pulse can be approximately estimated as \cite{morvan2021qutrit}
\begin{equation}
\varepsilon  = \frac{1}{12}(\frac{2}{T_{2}^{(01)}} + \frac{2}{T_{2}^{(12)}} + \frac{2}{T_{2}^{(02)}} +\frac{1}{T_{1}^{(01)}} +\frac{1}{T_{1}^{(12)}}) \tau ,
\end{equation}
where $\tau$ (with $\tau \ll T_{1},T_{2}$ )denotes the gate duration. Note that to be consistent with the random benchmarking (RB) results, we adopt the average gate error as the metric, which is related to the RB depolarization decay parameter by $(1-p)(1-1/3)$. Here, we estimate the incoherent error solely based on the Ramsey dephasing times reported in Table~\ref{Table_1}. The incoherent error due to decoherence processes is approximately $0.118 \times 10^{-3} \times (35+5) = 4.7 \times 10^{-3} $ for a 35-ns gate (IRB)
and $0.118 \times 10^{-3} \times (35+5) \times 1.667 = 7.868 \times 10^{-3} $
for an average Clifford gate (RB), where 35 ns is the duration of a native pulse and 5 ns the delay between two consecutive pulses. 

In our protocol, the dominant source of coherent error for single-qutrit gates arises from cross-coupling effects, where the applied drive inevitably off-resonantly couples to unintended transitions in addition to the target transition. This issue is particularly significant in our device, which exhibits the relatively weak anharmonicity of approximately 193 MHz in the transmon, resulting in substantial coherent errors. To comprehensively investigate the impact of cross-coupling effects on the infidelities of H and X gates, we perform the numerical simulations affected by coherent errors over a range of anharmonicities without applying additional calibrations.
In addition, we evaluate the dependence of coherent errors on gate duration, also in the absence of extra calibrations.

\renewcommand{\thefigure}{S\arabic{figure}} 
\begin{figure}[ht]
    \centering
    \includegraphics[width = 0.85\textwidth, height = 0.25\textheight]{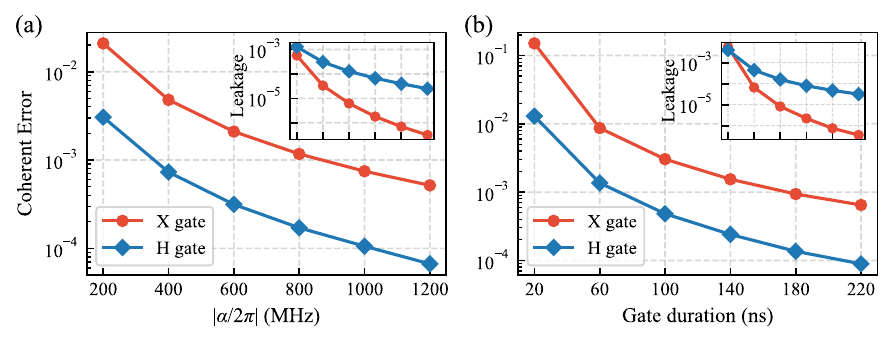}
    \caption{\justifying Coherent error budgets of the H gate and X gate as a function of (a) anharmonicity with a fixed gate duration of $T$ = 35 ns and (b) gate duration with anharmonicity fixed at $ -200$ MHz, without calibration. The insets show the separately extracted leakage errors for both gates. As the anharmonicity or gate duration increases, the coherent errors, including leakage, significantly decrease for both the H and X gates.}
    \label{fig:s8}
\end{figure}

As illustrated in Fig.~\ref{fig:s8}(a), an increase in the anharmonicity leads to a reduction in coherent errors, as the enhanced level spacing suppresses unwanted cross-coupling between transitions. Fig.~\ref{fig:s8}(b) further demonstrates that longer gate times also reduce coherent errors, since a longer gate requires a weaker drive amplitude, which in turn results in weaker off-resonant excitation and hence alleviating cross-coupling. Furthermore, we observe that the coherent error of the H gate is generally smaller than that of the X gate, primarily because the average Rabi frequency required to implement the H gate is lower.

The numerical results suggest that our gate scheme is particularly well suited for high-anharmonicity devices, such as capacitively shunted flux qutrits \cite{yurtalan2020implementation}, where gate infidelity is predominantly limited by decoherence. In contrast, for low-anharmonicity devices like the transmon qutrit used in our experiment, coherent errors play a more significant role. This leads to a trade-off between coherent and incoherent errors, indicating the need for systematic calibration to enable the optimal selection for gate times. Note that leakage errors have a negligible contribute to the total gate error for our proposed protocol, as illustrated in Fig.~\ref{fig:s8}. Consequently, we focus on mitigating phase and amplitude errors, which dominate the coherent errors in experiments, by calibrating control parameters such as pulse amplitudes, detunings, and relative phases.

\subsection{Calibration details}

To assess gate performance throughout the optimization process, we employ Clifford-based RB with a fixed sequence length of $m = 50$ \cite{kelly2014optimal}. This selection strikes a balance between experimental runtime and sensitivity to gate errors. In contrast to quantum process tomography (QPT), which can conflate intrinsic gate errors with state preparation and measurement (SPAM) errors, RB offers a more reliable estimate of the average error per gate through statistical averaging across numerous random sequences.

In our experiment, the single-qutrit Clifford group $\mathcal{C}_3$ is implemented using the gate set including H and X gates, requiring on average 
1.542 H and 0.125 X gates
per Clifford as stated in Section \ref{sec:C3gates}. Consequently, the measured Clifford error rate indeed reflects errors associated with both gates, particularly coherent contributions.
Therefore, during the optimization, control parameters are iteratively adjusted to minimize the observed Clifford error.
We use the Sequential Least Squares Programming (SLSQP) algorithm, which rapidly converges to a local minimum where the error per Clifford reaches its lowest observed value within the explored parameter space. At this point, the local minimum yields a set of optimal control parameters that represent the best experimental implementation of the H gate and X gate in our experimental setting.

In detail, we begin by analyzing the experimental parameters required for the X gate, viz.,
\begin{subequations}\label{eqs:40}
\begin{align}
\tilde{V}_{1}(t) &= [ A_1\Omega_{1}(t) + \frac{i\lambda_1 A_1}{\alpha}\dot{\Omega}_{1}(t) ]\cdot e^{-i(\omega_{01} + \Delta_1) t}, \\
\tilde{V}_{2}(t) &= [ A_2\Omega_{2}(t) + \frac{i\lambda_2 A_2}{\alpha}\dot{\Omega}_{2}(t) ]\cdot e^{-i(\omega_{12} + \Delta_2) t},
\end{align}
\end{subequations}
and the H gate 
\begin{subequations}\label{eqs:41}
\begin{align}
\tilde{V}_{1}(t) &=  A_1\Omega_{1}(t) \cos\left(\omega _{01}t  + \int_{0}^{t} 0.6581B_1\Omega_{1} (\tau)d\tau\right),  \\
\tilde{V}_{2}(t) &=  A_2\Omega_{2}(t) \cos\left(\omega _{12}t  - \int_{0}^{t} 0.6581B_2\Omega_{2} (\tau)d\tau\right) .
\end{align}
\end{subequations}
Here, $\tilde{V}_{1}(t)$ and $\tilde{V}_{2}(t)$ represent the microwave pulse envelopes, and  $A_1$ and $A_2$ establish the linear relationship between the applied driving voltages and the corresponding theoretical pulse amplitudes. To mitigate coherent errors, we introduce empirically calibrated DRAG corrections ($\lambda$) and frequency corrections ($\Delta$) for the X gate. However, for the H gate, only frequency correction was required, due to its comparatively weaker drive amplitude.

Note that the frequency correction approach for the H gate differs from that used for the X gate. This distinction stems from the implementation of the H gate via a chirped pulse, where the chirp frequency is linearly proportional to the pulse amplitude, with a ratio of 0.6581 (See Section~\ref{sec:1} for details).
We also need to compensate for the virtual phase gates in H and X gates, as demonstrated in Section~\ref{sec:1}. That is, $X = Vz_{post}(\phi_{x1}, \phi_{x2}) \cdot U_{x} \cdot Vz_{pre}(\phi_{x3}, \phi_{x4})$ and $H = U_{h} \cdot Vz_{pre}(\phi_{h1}, \phi_{h2})$.
Here, $Vz_{post}$ is omitted for the H gate as it induces no physical effect under our qutrit gate set.

\renewcommand{\arraystretch}{1.2} 
\begin{table}[htbp]
	\centering
    \caption{Gate parameters}
    \label{tab:Table_2}
    \setlength{\tabcolsep}{5mm}{
	\begin{tabular}{lllllll}
        \toprule
        \midrule
		X gate  & $A_{x1}$    & $A_{x2}$    & $\Delta_{x1}$ & $\Delta_{x2}$ & $\lambda_{x1}$ & $\lambda_{x2}$ \\
        \midrule
		H gate  & $A_{h1}$    & $A_{h2}$    & $B_{h1}$      & $B_{h2}$      & $\phi_{h1}$    & $\phi_{h2}$    \\
        \midrule
		$Vz$ gate & $\phi_{x1}$ & $\phi_{x2}$ & $\phi_{x3}$   & $\phi_{x4}$   &           &     \\
        \midrule
        \bottomrule
	\end{tabular}}
\end{table}

For simplicity, we employed distinct repetitive gate sequences: X gate sequences for parameters in the first row of Table~\ref{tab:Table_2}, H gate sequences for the second row, and HXH combinations for the third row.
This method systematically reduces the problem dimensionality.

\renewcommand{\thefigure}{S\arabic{figure}} 
\begin{figure}[ht]
    \centering
    \includegraphics[width = 0.85\textwidth, height = 0.28\textheight]{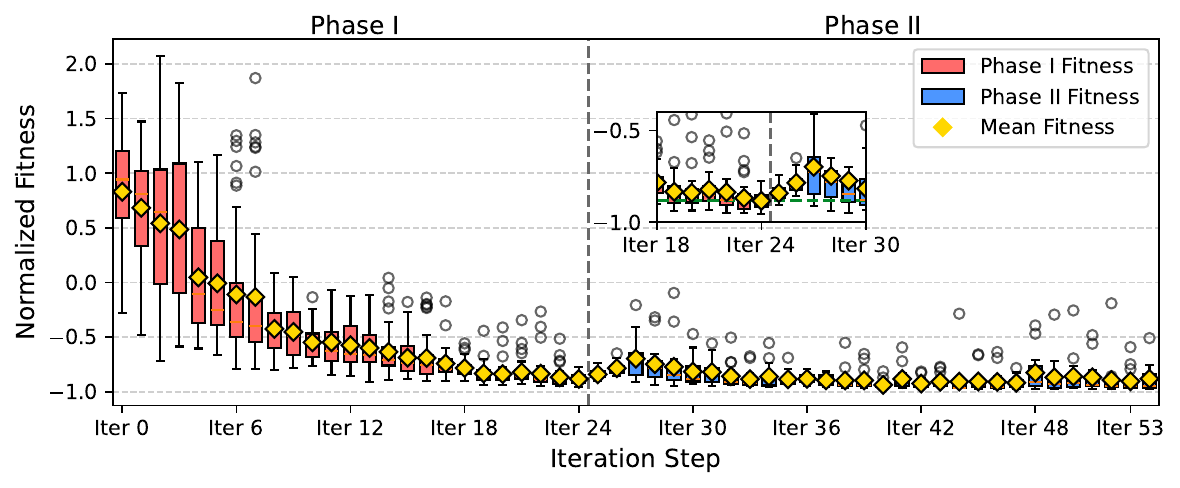}
    \caption{
    Normalized fitness convergence under dual-phase optimization is illustrated in two phases. In Phase I (red), with high mutation and crossover rates, the optimization proceeds across 40 initial Latin Hypercube Sampling (LHS) populations and 5 training RB sequences, reaching a convergence threshold of 88\%.
    Phase II (blue) is initialized with the results from Phase I and employs lower mutation and crossover rates, evaluated on 6 unseen RB sequences (inset: Iterations 18–30).
    The mean fitness trajectories (yellow diamonds) reveal rapid stabilization during Phase I and exhibit ultra-low variation ($<4.8\%$) during Phase II validation, demonstrating effective suppression of overfitting.
    These results are comparable to the generalization capabilities of machine learning models applied to unseen data
    }
    \label{fig:s50}
\end{figure}

However, optimizing only four or six parameters with repetitive sequences frequently results in overfit solutions.
Furthermore, the quantum process tomography as the fidelity metric for calibration is also unsuitable, given that it is tendency to introduce additional readout errors and to cap the achievable qutrit gate fidelity at a level constrained by single-qubit gate errors.
Quantum process tomography significantly increases computational complexity and imposes excessive demands on system stability.
Routinely, applying Randomized Benchmarking (RB) directly to calibrate experimental parameters is a simple and robust way to yield the optimal performance.

Next, we initialize 40 distinct population sets via Latin hypercube sampling (LHS), adopting elevated mutation and crossover rates to prevent convergence into local minima.
To mitigate overfitting, five different RB sequences were generated, with the objective function $Z$ computed using
\begin{subequations}\label{eqs:43}
    \begin{align}
    &f(x, p) = A\cdot p^x + B,  \\
    &Z = \frac{1}{N}\sum_{j=1}^{N} \left( 0.3\cdot \epsilon - p_{\text{fit}} \right).
    \end{align}
\end{subequations}
Here, $f(x, p)$ represents the fitted RB sequence curve, $\epsilon$ denotes the fitting residual error and $p_{\text{fit}}$ is the fitted gate fidelity.
Upon reaching 88\% convergence in Phase I, early stopping was triggered.
Subsequently, six new RB sequences were generated, and the objective function was reevaluated under reduced mutation and crossover rates.
The Phase I final population then initialized Phase II, enabling expanded parameter space exploration until near-complete convergence.
This two-phase strategy effectively suppresses overfitting while accelerating convergence efficiency.
As shown in Fig.~\ref{fig:s50}, the objective function and critical parameters exhibit consistent convergence trajectories. Crucially, Phase I optimizes parameters maintained high stability ($<4.8\%$ variation) during Phase II validation, confirming robustness against independent sequence sets. This performance consistency, analogous to an ML model generalizing successfully to unseen validation data, demonstrates effective overfitting mitigation throughout the calibration workflow.


Consequently, we perform population dynamics measurements (Fig. 2 in main text) and interleaved RB (Fig. 3 in main text).
The results independently validate the optimized gate performance and ensure consistency across calibration methods, confirming the suppression of coherent errors and provide further insight into the dominant error mechanisms.

\clearpage

\makeatletter
\let\old@algocf@capt@plain\algocf@capt@plain
\let\algocf@capt@plain\@undefined
\makeatother

\begin{algorithm}[H]
    \centering
    \begin{minipage}{0.9\linewidth}
    \caption{Construction of SU(3) Clifford Group from Generators}
    \label{alg:su3_clifford}
    \SetAlgoLined
    \KwIn{Qudit dimension $d$ (default: 3 for SU(3))}
    \KwOut{Complete Clifford group for SU($d$)}
    
    \SetKwProg{Fn}{Function}{:}{end}
    
    \Fn{GenerateFundamentalGates($d$)}{
        $\omega \gets e^{2\pi i / d}$ \tcp*{Primitive root of unity}
        $Z \gets \text{diag}(\omega^0, \omega^1, \omega^2)$ \tcp*{Generalized Z gate}
        $S \gets \text{diag}(\omega^0, \omega^1, \omega^3)$ \tcp*{Generalized S gate}
        $X \gets \text{cyclic shift matrix}$ \tcp*{Generalized X gate}
        $H \gets \frac{1}{\sqrt{d}}[\omega^{ij}]$ \tcp*{Generalized H gate}
        \Return {$[Z, S, X, H]$} \tcp*{Return the 4 generators}
    }
    
    \Fn{TryAddProducts($\text{Group}, h, k$)}{
        \textbf{Compute product} $g_1 \gets \text{Group}[h] \times \text{Group}[k]$\;
        \textbf{Compute product} $g_2 \gets \text{Group}[k] \times \text{Group}[h]$\;
        $added \gets \text{False}$\;
        \If{$g_1$ not in Group}{
            $\text{Group} \gets \text{Group} \cup \{g_1\}$\;
            $added \gets \text{True}$\;
        }
        \If{$g_2$ not in Group}{
            $\text{Group} \gets \text{Group} \cup \{g_2\}$\;
            $added \gets \text{True}$\;
        }
        \Return {$\text{Group}, added$}\;
    }
    
    \BlankLine
    \textbf{Main Algorithm:}\;
    $N \gets 216$ \tcp*{Theoretical size of SU(3) Clifford group}
    $\text{Group} \gets \text{GenerateFundamentalGates}(d)$ \tcp*{Initialize with generators}
    
    $h \gets 0, j \gets 0, L \gets \text{len}(\text{Group})$\;
    \While{$h < L$}{
        $addedNew \gets \text{False}$\;
        \For{$k \gets j$ \KwTo $L-1$}{
            $\text{Group}, added \gets \text{TryAddProducts}(\text{Group}, h, k)$\;
            \If{added}{
                $addedNew \gets \text{True}$\;
            }
        }
        \uIf{$\text{len}(\text{Group}) = N$}{
            \textbf{break} \tcp*{Group is complete}
        }
        \uElseIf{not addedNew}{
            $h \gets h + 1$ \tcp*{Move to next starting element}
            $j \gets 0$\;
        }
        \Else{
            $j \gets L$ \tcp*{Continue with new elements}
            $L \gets \text{len}(\text{Group})$\;
        }
    }
    \Return $\text{Group}$\;
    \end{minipage}
\end{algorithm}

\begin{algorithm}[H]
    \centering
    \begin{minipage}{0.9\linewidth}
    \caption{\justifying Decomposition of SU(3) Matrix and Fidelity Calculation}
    \label{alg:su3_decomposition}

    \SetAlgoLined
    \KwIn{$U_{\text{target}}$ (optional), an SU(3) matrix. If not provided, generate randomly.}
    \KwOut{Phase parameters for 5 phase gates $(a_i, b_i)$, and fidelity value.}
    
    \eIf{$U_{\text{target}} == \text{None}$}{
        $U_{\text{target}} \gets \text{random\_su3}()$ \tcp*{Generate random SU(3) matrix}
    }{
        \tcp*{Use provided matrix}
    }

    $H \gets \text{su3\_h\_gate}()$ \tcp*{generalized Hadamard}

    \textbf{Initialize} bounds $\gets [(0, 2\pi)] \times 10$ \tcp*{Parameter bounds for optimization}

    \SetKwProg{Fn}{Function}{:}{end}
    
    \Fn{objective\_function(params, target\_u, H)}{
        $a_1, b_1, a_2, b_2, a_3, b_3, a_4, b_4, a_5, b_5 \gets params$\;
        $D_1 \gets \text{su3\_phase\_gate}(a_1, b_1)$\;
        $D_2 \gets \text{su3\_phase\_gate}(a_2, b_2)$\;
        $D_3 \gets \text{su3\_phase\_gate}(a_3, b_3)$\;
        $D_4 \gets \text{su3\_phase\_gate}(a_4, b_4)$\;
        $D_5 \gets \text{su3\_phase\_gate}(a_5, b_5)$\;
        $U_{\text{approx}} \gets D_1 H D_2 H D_3 H D_4 H D_5$\;
        $\text{fidelity} \gets |\frac{1}{3}\text{Tr}(U_{\text{approx}}^\dagger U_{\text{target}})|^2$\;
        $\text{Infidelity} \gets |1-\text{fidelity}|$\;
        \Return {Infidelity} \tcp*{Return the infidelity for minimization}
    }

    Result $\gets$ differential\_evolution(objective\_function, bounds, args=($U_{\text{target}}$, H)) \tcp*{Optimization Algorithm}
    phase\_params $\gets$ reshape(result.x, (5, 2)) \tcp*{Reshape to 5×2 parameter matrix}

    \textbf{Reconstruct} matrix using optimal parameters:\;
    $D_1 \gets \text{su3\_phase\_gate}(\text{phase\_params}[0])$\;
    $D_2 \gets \text{su3\_phase\_gate}(\text{phase\_params}[1])$\;
    $D_3 \gets \text{su3\_phase\_gate}(\text{phase\_params}[2])$\;
    $D_4 \gets \text{su3\_phase\_gate}(\text{phase\_params}[3])$\;
    $D_5 \gets \text{su3\_phase\_gate}(\text{phase\_params}[4])$\;
    $U_{\text{reconstructed}} \gets D_1 H D_2 H D_3 H D_4 H D_5$\;

    $\text{final\_fidelity} \gets |\frac{1}{3}\text{Tr}(U_{\text{reconstructed}}^\dagger U_{\text{target}})|^2$ \tcp*{Calculate final fidelity}

    \Return {phase\_params, final\_fidelity} \tcp*{Return optimal parameters and achieved fidelity}
    \end{minipage}
\end{algorithm}

~\\
~\\
~\\

%